\documentclass[prb,showpacs,twocolumn,aps,superscriptaddress,a4paper]{revtex4}
\usepackage{dcolumn}
\usepackage{amsmath}
\usepackage{graphicx}
\usepackage{latexsym}  
\usepackage{amsfonts}  
\usepackage{amssymb} 
\usepackage{color}
\DeclareGraphicsExtensions{.eps,.pdf,.gif,.jpg,.png}  
  
\newcommand{\be}{\begin{equation}}  
\newcommand{\ee}{\end{equation}}
\newcommand{\beq}{\begin{eqnarray}}  
\newcommand{\eeq}{\end{eqnarray}}  
   
\newcommand{\hH}{\hat{H}}

\begin{document}  
      
\def\bbe{\mbox{\boldmath $e$}}  
\def\bbf{\mbox{\boldmath $f$}}      
\def\bg{\mbox{\boldmath $g$}}  
\def\bh{\mbox{\boldmath $h$}}  
\def\bj{\mbox{\boldmath $j$}}  
\def\bq{\mbox{\boldmath $q$}}  
\def\bp{\mbox{\boldmath $p$}}  
\def\br{\mbox{\boldmath $r$}}      
  
\def\bone{\mbox{\boldmath $1$}}      
  
\def\dr{{\rm d}}  
  
\def\tb{\bar{t}}  
\def\zb{\bar{z}}  
  
\def\tgb{\bar{\tau}}  
        
\def\bC{\mbox{\boldmath $C$}}  
\def\bG{\mbox{\boldmath $G$}}  
\def\bH{\mbox{\boldmath $H$}}  
\def\bK{\mbox{\boldmath $K$}}  
\def\bM{\mbox{\boldmath $M$}}  
\def\bN{\mbox{\boldmath $N$}}  
\def\bO{\mbox{\boldmath $O$}}  
\def\bQ{\mbox{\boldmath $Q$}}  
\def\bR{\mbox{\boldmath $R$}}  
\def\bS{\mbox{\boldmath $S$}}  
\def\bT{\mbox{\boldmath $T$}}  
\def\bU{\mbox{\boldmath $U$}}  
\def\bV{\mbox{\boldmath $V$}}  
\def\bZ{\mbox{\boldmath $Z$}}  
  
\def\bcalS{\mbox{\boldmath $\mathcal{S}$}}  
\def\bcalG{\mbox{\boldmath $\mathcal{G}$}}  
\def\bcalE{\mbox{\boldmath $\mathcal{E}$}}  
  
\def\bgG{\mbox{\boldmath $\Gamma$}}  
\def\bgL{\mbox{\boldmath $\Lambda$}}  
\def\bgS{\mbox{\boldmath $\Sigma$}}  
  
\def\bgr{\mbox{\boldmath $\rho$}}  
  
\def\a{\alpha}  
\def\b{\beta}  
\def\g{\gamma}  
\def\G{\Gamma}  
\def\d{\delta}  
\def\D{\Delta}  
\def\e{\epsilon}  
\def\ve{\varepsilon}  
\def\z{\zeta}  
\def\h{\eta}  
\def\th{\theta}  
\def\k{\kappa}  
\def\l{\lambda}  
\def\L{\Lambda}  
\def\m{\mu}  
\def\n{\nu}  
\def\x{\xi}  
\def\X{\Xi}  
\def\p{\pi}  
\def\P{\Pi}  
\def\r{\rho}  
\def\s{\sigma}  
\def\S{\Sigma}  
\def\t{\tau}  
\def\f{\phi}  
\def\vf{\varphi}  
\def\F{\Phi}  
\def\c{\chi}  
\def\w{\omega}  
\def\W{\Omega}  
\def\Q{\Psi}  
\def\q{\psi}  
  
\def\ua{\uparrow}  
\def\da{\downarrow}  
\def\de{\partial}  
\def\inf{\infty}  
\def\ra{\rightarrow}  
\def\bra{\langle}  
\def\ket{\rangle}  
\def\grad{\mbox{\boldmath $\nabla$}}  
\def\Tr{{\rm Tr}}  
\def\Re{{\rm Re}}  
\def\Im{{\rm Im}}

\def\mol{{N_{\mathrm{mol}}}}
\def\molhat{{\hat{N}_{\mathrm{mol}}}}
\def\ud{{\mathrm{d}}}

\title{Image charge dynamics in time-dependent quantum transport}

\author{Petri My\"oh\"anen}

\author{Riku Tuovinen}

\author{Topi Korhonen}
\affiliation{Department of Physics, Nanoscience Center, FIN 40014, University of Jyv\"askyl\"a,Jyv\"askyl\"a, Finland}

\author{Gianluca Stefanucci}
\affiliation{Dipartimento di Fisica, Universit\`a di Roma Tor Vergata, Via della Ricerca Scientifica 1, I-00133 Rome, Italy}
\affiliation{Laboratori Nazionali di Frascati, Istituto Nazionale di Fisica
Nucleare, Via E. Fermi 40, 00044 Frascati, Italy}
\affiliation{European Theoretical Spectroscopy Facility (ETSF)}

\author{Robert van Leeuwen}
\affiliation{Department of Physics, Nanoscience Center, FIN 40014, University of Jyv\"askyl\"a,Jyv\"askyl\"a, Finland}
\affiliation{European Theoretical Spectroscopy Facility (ETSF)}

\date{\today}  
\begin{abstract}
In this work we investigate the effects of the electron-electron 
interaction between a molecular junction and the metallic leads in 
time-dependent quantum transport. We employ the recently developed 
embedded Kadanoff-Baym method [Phys. Rev. B {\bf 80}, 115107 (2009)] 
and show that the molecule-lead interaction changes substantially  
the transient and steady-state transport properties. 
We first show that the mean-field Hartree-Fock (HF) approximation 
does not capture the polarization effects responsible for the 
renormalization of the molecular levels   neither in 
nor out  of equilibrium. Furthermore, due to the time-local nature of the HF 
self-energy there exists a region in parameter space for which the system 
does not relax after the switch-on of a bias voltage. These and other 
artifacts of the HF approximation disappear when including 
correlations at the second-Born or GW levels. Both these 
approximations contain polarization diagrams which correctly account for the 
screening of the charged molecule. We find that by changing the 
molecule-lead interaction the ratio between the screening and relaxation time changes,
an effect which must be properly taken into 
account in any realistic time-dependent simulation. Another important 
finding is that while in equilibrium the molecule-lead interaction is 
responsible for a reduction of the HOMO-LUMO gap and for a substantial 
redistribution of the spectral weight between the main spectral peaks and 
the induced satellite spectrum, in the biased system it can have 
the opposite effect, i.e., it {\em sharpens} the 
spectral peaks and {\em opens} the HOMO-LUMO gap.
\end{abstract}
  
  
\maketitle  

\section{Introduction}
The electron transport through molecular devices has gained 
remarkable interest during last years, primarily 
due to experimental advances in creating conductive molecule-metal 
junctions.\cite{reed,smit} From the experimental point 
of view these systems are very attractive for their potential 
utilization as the next-generation nanometer scale building blocks  
for future integrated circuits exceeding up to terahertz operating 
frequencies. For theorists, the experimental realization of  
electron transport through molecules opens up a new intriguing and 
challenging playground for both theoretical and numerical  
modelling of the underlying physical processes. Understanding these 
processes at a microscopic level is crucial for the future 
development of  \emph{molecular electronics}. 

Considerable progress has been made to investigate both
steady-state\cite{Meir.1992,taylor.2001,brand.2002,rocha.2006,Cardamone.2009,darencet.2007,
kr3.2008,Bergfield.2009,ness.2010,cornean.2011,knap.2001} and 
time-dependent\cite{Jauho.1994,GS.2004,Myohanen.2008,Myohanen.2009,Moldoveanu.2009,
Tomita.2009,Moldoveanu.2010,baer.2004,GS.2005,zheng.2007,bokes.2008,pump.2008,
Zheng.2010,marc.2010,Evans.2009,varga.2001,cohen.2011,cuansing} 
transport properties of metal-nanostructure-metal junctions. As an 
increasing trend the system is partitioned into an 
explicitely treated interacting region coupled to   
noninteracting electron reservoirs (leads) which act as source and sink  
terminals. However, the partitioning into an interacting
and a noninteracting part is, in general, not well justified 
due to the long range nature of the Coulomb interaction.
Recently, there have been some advances in calculating 
transport properties of nanoscale junctions while incorporating the electron-elecron 
interaction in the leads. Perfetto et al.\cite{perf} recently found that modelling the 
electron-electron interaction in low-dimensional leads with 
the Luttinger model the initial correlation 
effects are not washed out in the long-time limit and contribute 
substantially to the steady-state current.
Bohr \emph{et al}.\cite{bohr1} and Borda \emph{et al}.\cite{Borda} 
investigated the effects of the lead-molecule interactions 
in the interacting resonant level model
and showed that it can lead to a strong enhancement of the conductance.
More recently these studies have been extended to 
long-range lead-molecule interactions\cite{Elste, perf2}.\\
Considerable attention has also been devoted to the effects of surface 
polarization (or image charge formation). In Refs. 
\onlinecite{Neaton, kr1.2009, kr2.2009, Kaasbjerg1,Kaasbjerg2} it was shown that polarization 
effects can dramatically change the 
quasi-particle gap of molecules near the metallic surfaces where the dynamical correlation effects and 
molecule-lead hopping integrals reduce the molecular energy gap across the binding regime from gas phase 
to physisorption. Clearly, this renormalization of the molecular levels 
can have a large impact on the  
transport properties of weakly coupled molecular junctions. 
Yet, the question of how 
the molecule-lead interactions and, consequently, 
the formation of an image charge affects the ultrafast electron 
dynamics before a steady-state (if any) is reached  is still 
unanswered. The present paper wants to address two fundamental 
issues: what is the time-scale to screen  molecular charge 
fluctations induced by the sudden switch-on of an external bias? And what are 
the scattering processes (or Feynman diagrams) relevant for an 
accurate description of the screening {\em and} relaxation dynamics? 

To answer these questions we will use the Kadanoff-Baym method
which has recently been applied to both finite isolated \cite{Dahlen, Balzer1, Balzer2, Marc1, Marc2} and quantum transport systems 
\cite{Myohanen.2008, Myohanen.2009,Myohanen.2010} and 
has the merit of preserving all basic conservation 
laws \cite{Kadanoff,Baym.1962}. We show that the mean-field 
Hartree-Fock approximation suffers from several limitations in this 
context. Besides being unable to account for dynamical polarization 
effects the Hartree-Fock approximation can give rise to ``unstable''
time-dependent solutions  with persistent 
oscillations in density and current. All mean-field artifacts 
disappear when including polarization effects in the self-energy, 
either at the second-Born or GW level. These correlated solutions 
have recently been assessed in the Anderson 
model\cite{anna-maija} and good agreement with time-dependent Density Matrix 
Renormalization Group (DMRG) 
data was found.\cite{feiguin} Here we employ them for a thorough 
analysis of the screening versus relaxation dynamics as a function of 
the interaction strength, the molecule-lead hopping integrals and 
the external bias.
We find that the relaxation time $\t_{\rm rel}$
becomes shorter when increasing the
molecule-lead interactions at second-Born and GW level while the screening 
time $\tau_{\rm scr}$ is roughly independent on the interaction strength. 
Often, the time-dependent quantum transport 
simulations are based on the assumption that $\t_{\rm scr}/\t_{\rm 
rel}\ll 1$. Our results show that the molecule-lead 
interaction can substantially increase this ratio.
Another remarkable effect of the molecule-lead interaction is that
for large enough biases the electronic correlations can {\em sharpen} the 
spectral peaks and {\em widen} the gap between the levels of Highest Occupied
Molecular Orbital (HOMO) and Lowest Unoccupied Molecular Orbital (LUMO). This behavior is 
exactly the opposite of the equilibrium behavior and indicates that 
in the presence of a current flow the screening lenghtens the 
HOMO-LUMO quasi-particle life-time and decreases (increases) the  
ionization potential (electron affinity).\\
\\
The article is organized as follows.
In Section \ref{mod+the} we introduce the model Hamiltonian for quantum 
transport simulations and discuss the exact solution for zero molecule-lead hopping 
integrals.  
We also give a short account of the theoretical background and defer 
the reader to previously published work for details.
In Section \ref{competing} we analyze the screening versus relaxation 
time and the effect of the formation of an image charge in the 
equilibrium spectral function. Section \ref{sh-time-dyn} deals with 
the short-time dynamics of the lead-molecule-lead junction driven 
out of equilibrium by the sudden switch-on of a constant bias while 
Section \ref{long-tim-dyn} deals with the long-time dynamics, and in 
particular with the absence of relaxation within HF and the  effects of 
screening in the $I-V$ characteristic. 
The main conclusions are then drawn in Section \ref{conc}.
 
\section{Image charge model}
\label{mod+the}

\subsection{Hamiltonian}

\begin{figure}[t]
\centering
\includegraphics[scale=0.5]{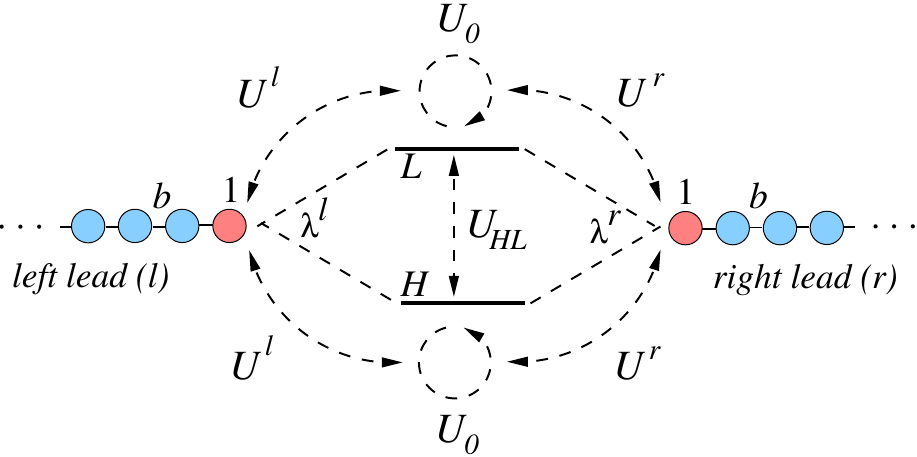}
\caption{Image charge model for quantum transport.}
\label{Fig1}
\end{figure}

To study the image charge effect we consider a model Hamiltonian
that was introduced in Refs. \cite{kr1.2009,kr2.2009}. 
This image charge model Hamiltonian is displayed 
schematically in Fig.~\ref{Fig1}
\beq\label{Hamiltonian}
\hat{H}(t) = \hat{H}_{\textnormal{mol}} + \hat{H}_{\textnormal{ch}}(t) + \hat{V} - \mu\hat{N}.
\eeq
The molecular region is modelled by a two-level system representing 
the Highest Occupied Molecular Orbital ($H$) and the Lowest 
Unoccupied Molecular Orbital ($L$) with energies 
$\epsilon_H$ and $\epsilon_L$ respectively
\beq\label{molecularham}
\hat{H}_{\textnormal{mol}} &=& \epsilon_H\hat{n}_H + 
\epsilon_L\hat{n}_L  
\nonumber \\
&+&
U_0(\hat{n}_{H\uparrow}\hat{n}_{H\downarrow} + 
\hat{n}_{L\uparrow}\hat{n}_{L\downarrow}) + U_{HL}\hat{n}_{H}\hat{n}_{L}, 
\eeq
The interaction strengths $U_0$ and $U_{HL}$ account for the 
intra-level and inter-level electron repulsion. 
Furthermore, we used the standard notation $\hat{n}_{i} = 
\sum_{\sigma=\uparrow\downarrow}\hat{n}_{i\sigma}$  
for the particle number operator of the molecular level $i=H,L$, where 
$\hat{n}_{i\sigma} = \hat{c}_{ i\sigma}^{\dagger}\hat{c}_{
i\sigma}$ and 
$\hat{c}_{i\sigma}^{\dagger}$ and $\hat{c}_{i\sigma}$ are the 
electron creation and annihilation operators.

The second term in Eq.~(\ref{Hamiltonian}) describes the left 
($\alpha=l$) and right ($\alpha=r$) leads,
\be
\hat{H}_{\textnormal{ch}}(t) = \sum_{\alpha=l,r} \sum_{i,j=1} 
\sum_{\sigma=\uparrow\downarrow}  
[ h_{ij}^{\alpha} + \delta_{ij}W^{\alpha}(t)]\hat{c}_{\alpha 
i\sigma}^{\dagger}\hat{c}_{\alpha j\sigma}, 
\label{hamch}
\ee
which are modelled as 
one-dimensional semi-infinite tight-binding (TB) chains subject to 
time-dependent  
uniform bias voltages $W^{\alpha}(t)$. The 
TB parameters $h_{ij}$ of the chain are chosen so that
$h_{ij} = b$ for $i,j$ nearest neighbours and zero otherwise. 
Finally, $\hat{c}_{\alpha i\sigma}^{\dagger}$ and $\hat{c}_{\alpha 
j\sigma}$   
are the creation and annihilation operators for electrons 
in lead $\a$, site $i=1,2,\ldots$ and spin 
$\sigma$.

The third term 
in Eq.~(\ref{Hamiltonian}) describes the interaction between the 
molecular levels and  the TB  chains
\begin{eqnarray}\label{interaction}
\hat{V} &=& 
\sum_{\alpha=l,r}\sum_{i=H,L}\sum_{\sigma=\uparrow\downarrow}
\lambda^{\alpha}(\hat{c}_{\alpha1\sigma}^{\dagger}\hat{c}_{ i\sigma} +    
\hat{c}_{ i\sigma}^{\dagger}\hat{c}_{\alpha 1 \sigma}) \nonumber\\ 
&& + \sum_{\alpha=l,r}   U^{\a}(\hat{n}_{\alpha 1}-1)(\molhat-2).
\end{eqnarray}
Here $\lambda^{\alpha}$ and $  U^{\a}$ are the hopping integrals 
(proportional to the hybridization of the molecular levels) and Coulomb 
interaction strengths between the HOMO/LUMO levels  
and the terminal site of lead $\alpha$. The quantity
$\hat{n}_{\alpha 1}$ is the particle number 
operator of site 1 of lead $\alpha$, $\hat{n}_{\alpha 1} = 
\sum_{\sigma=\uparrow\downarrow}\hat{c}_{\alpha1\sigma}^{\dagger}\hat{c}_{\alpha1\sigma}$, while 
$\molhat$ is
the total number of particle operator of the molecule,
$\molhat = \hat{n}_H + \hat{n}_L$. 
We consider the system initially in equilibrium at zero temperature, 
zero bias, $W^{\alpha}=0$, and at half-filling. Then, the average density on the 
lead sites is unity while the average density of the HOMO and LUMO 
levels is 2 and 0 respectively. To guarantee the charge neutrality of 
the interacting region we subtracted a 
positive background charge of 1 from $\hat{n}_{\alpha 1}$ and of $2$ 
from $\molhat$.

This complete the explanation and justification of the image charge 
model (ICM). It can be considered as an extension of the interacting resonant level 
model to study molecular excitons and polarization 
effects. The ICM can, of course, be further refined by including 
interactions in the leads and a direct lead-lead interaction, and 
can be further generalized to
two- or three-dimensional leads, more molecular levels, etc. 
Equation (\ref{Hamiltonian}), however, provides the minimal model to 
 study the effects of image charges in the 
non-equilibrium properties of nanoscale junctions and in this paper 
we will not discuss any of the aforementioned extensions.

\subsection{Uncontacted case: Exact solution}
\label{exsol}

The ICM can be solved exactly for zero hybridization, 
i.e., $\lambda^r=\lambda^l=0$. In this case the operators $\hat{n}_{H}$ and 
$\hat{n}_{L}$  commute with the Hamiltonian and hence
the number of electrons on the $H$ and $L$ levels are  
conserved quantities. Let us consider for simplicity the unperturbed 
Hamiltonian $\hat{H}$ obtained from Eq. (\ref{Hamiltonian}) by 
setting the bias $W^{\alpha}$ to zero.
All eigenstates of $\hat{H}$ have the form
\be\label{eigenstates}
| M,s\rangle = \prod_{j} \theta_j \hat{c}_j^{\dagger} |\Phi_s\rangle .
\ee
Here the $\hat{c}_j^{\dagger}$-operators create
electrons on the molecular level $j\in\{H\uparrow,H\downarrow,L\uparrow,L\downarrow\}$ 
and $\theta_j$ is either equal to one or zero depending on what states one likes to occupy.
The corresponding molecular 
configuration is specified by the collective quantum number $M$. 
The state $|\Phi_s\rangle$ is the $s$-th excited state of the uncontacted 
leads and has the property
 $\hat{n}_j|\Phi_s\rangle = 0$. For example, a state
with two electrons in the HOMO-level of the molecule is 
$|H\uparrow,H\downarrow,s \rangle = c_{H\uparrow}^{\dagger}c_{H\downarrow}^{\dagger}|\Phi_s\rangle$. 
To find the secular equation for the $|\Phi_s\rangle$ we apply 
$\hat{H}$ to $|M,s\rangle$ and find
\beq\label{singleleadsolutions}
&&\hat{H} | M,s\rangle \nonumber \\
&&=\left[\hat{H}_{\textnormal{mol}} + \hat{H}_{\textnormal{ch}} +  
\sum_{\a}U^{\a}  (\hat{n}_{\a 1}-1)(\molhat-2) \right]| M,s\rangle \nonumber \\ 
&&=\left[E_{M} + \mathcal{E}_{M,s}\right]| M,s\rangle \nonumber \\
&&=E_{M,s}|M,s\rangle,
\eeq
where $E_{M}$ is the total energy of the \emph{isolated} molecule 
with $N_{\rm mol}$ electrons satifying the eigenvalue equation 
\beq
\hat{H}_{\textnormal{mol}}|M,s\rangle = E_{M} |M,s\rangle,
\eeq
while $\mathcal{E}_{M,s}$ is the total energy of the {\em 
uncontacted} leads in the 
presence of the potential $U^{\a}  (N_{\rm mol}-2)$ at the terminal sites
\beq\label{nonintchain}
&&\hat{H}_{\textnormal{ch}} (U)| M,s\rangle \nonumber\\
&&\equiv 
\left[\hat{H}_{\textnormal{ch}}(U=0) +  \sum_{\a}U^{\a}  
(N_{\rm mol}-2)(\hat{n}_{\a 1}-1)\right]| M,s\rangle\nonumber \\ 
&&= \mathcal{E}_{M,s}| M,s\rangle.
\eeq
This potential depends on the strength of the Coulomb 
interaction $ U^{\a}  $ and on the number  $N_{\rm mol}$ of electrons on the 
molecule. Once we know the electronic 
configuration of the molecule, the problem reduces to solving 
the eigenvalue equation (\ref{nonintchain}) for a noninteracting TB 
chain with an impurity-like potential at the terminal site.  If the 
molecule is charge-neutral, $N_{\rm mol}=2$, this  
potential is zero. However, adding  (removing) an 
electron from the charge-neutral molecule gives rise to a potential 
$+U^{\a}$  ($-U^{\a}$). 
 This, in turn, causes a 
depletion/accumulation of charge which is exactly the image charge.

It is worth stressing that  the presence of the lead-molecule 
interaction  affects the total energies of the charged system, 
see again Eq. (\ref{nonintchain}),  and  
consequently changes the addition and removal energies.
Consider, for instance, the solution for  
a simple 2-site chain and a lead-molecule interaction $U^{r}=U$ and 
$U^{l}=0$ (no coupling to the left lead).
It is easy to show that  
the electron affinity is  $A = \epsilon_L + 2U_{HL} + 2|b| - 2\sqrt{( U  /2)^2 + b^2}$ 
while the ionization energy is 
$I = \epsilon_H + U_0 - 2|b| + 2\sqrt{( U  /2)^2 + b^2}$  (see Appendix 
\ref{2sitesol}).  The 
difference $A-I$ reduces with increasing $U$ and the quasi-particle 
gap collapses.
This can also be viewed from another, more general, point of view.
Consider for simplicity that $U^{\a}=U$ for both leads and that the 
intra-molecular interactions $U_0$ and $U_{HL}$ are zero. 
If the molecule is charge neutral ($N_{\rm mol}=2$) the energies of the $N$ and 
$N \pm 1$ particle ground states (with the constraint that the electron is added to or removed 
from the molecule) are given by
\beq
E_N &=& 2 \epsilon_H + \mathcal{E}_{\rm GS} (0) \\
E_{N+1} &=& 2 \epsilon_H + \epsilon_L + \mathcal{E}_{\rm GS} (U) \\
E_{N-1} &=& \epsilon_H + \mathcal{E}_{\rm GS} (-U) 
\eeq
where we defined $\mathcal{E}_{\rm GS} (U)$ to be the ground state 
energy of the Hamiltonian $\hat{H}_{\rm ch} (U)$ of Eq. (\ref{nonintchain}).
Therefore, the electron affinity $A$ and ionization energy $I$
read
\beq
A= E_{N+1} - E_{N} = \epsilon_L + \mathcal{E}_{\rm GS} (U) - \mathcal{E}_{\rm GS} (0) 
\label{aff}\\
I= E_{N} - E_{N-1} = \epsilon_H + \mathcal{E}_{\rm GS} (0) - \mathcal{E}_{\rm GS} (-U) .
\label{ion}
\eeq
Let $|\Phi_{\rm GS}(u)\ket$ be the ground state of $\hat{H}_{\rm ch} 
(U)$. Then, according to Hellman-Feynman theorem \cite{Feynman} 
\be
\frac{d  \mathcal{E}_{\rm GS} (u)}{d u} = \langle \Phi_{\rm GS} (u) | 
\frac{d\hat{H}_{\rm ch} (u)}{du} | \Phi_{\rm GS} (u) \rangle, 
\ee
and therefore
\be\label{HellmanFeynman}
\mathcal{E}_{\rm GS}( U  ) - \mathcal{E}_{\rm GS}(0) =  \sum_{\alpha} 
\int_0^{ U  }[ n_{\alpha 1} (u)-1]\ud u. 
\ee
 
From this equation we see clearly how the ground state energy depends 
on the molecular occupation: 
If we add an electron to the molecule we push away charge from the 
first sites of the leads and hence 
the integral is negative and the affinity lowers.
On the other hand, if we remove an electron from the molecule we 
attract charge to the first sites of the leads and the 
ionization energy increases. \\
\\
\begin{figure}[t]
\centering
\includegraphics[trim = 19mm 00mm 00mm 0mm, clip, width=0.49\textwidth]{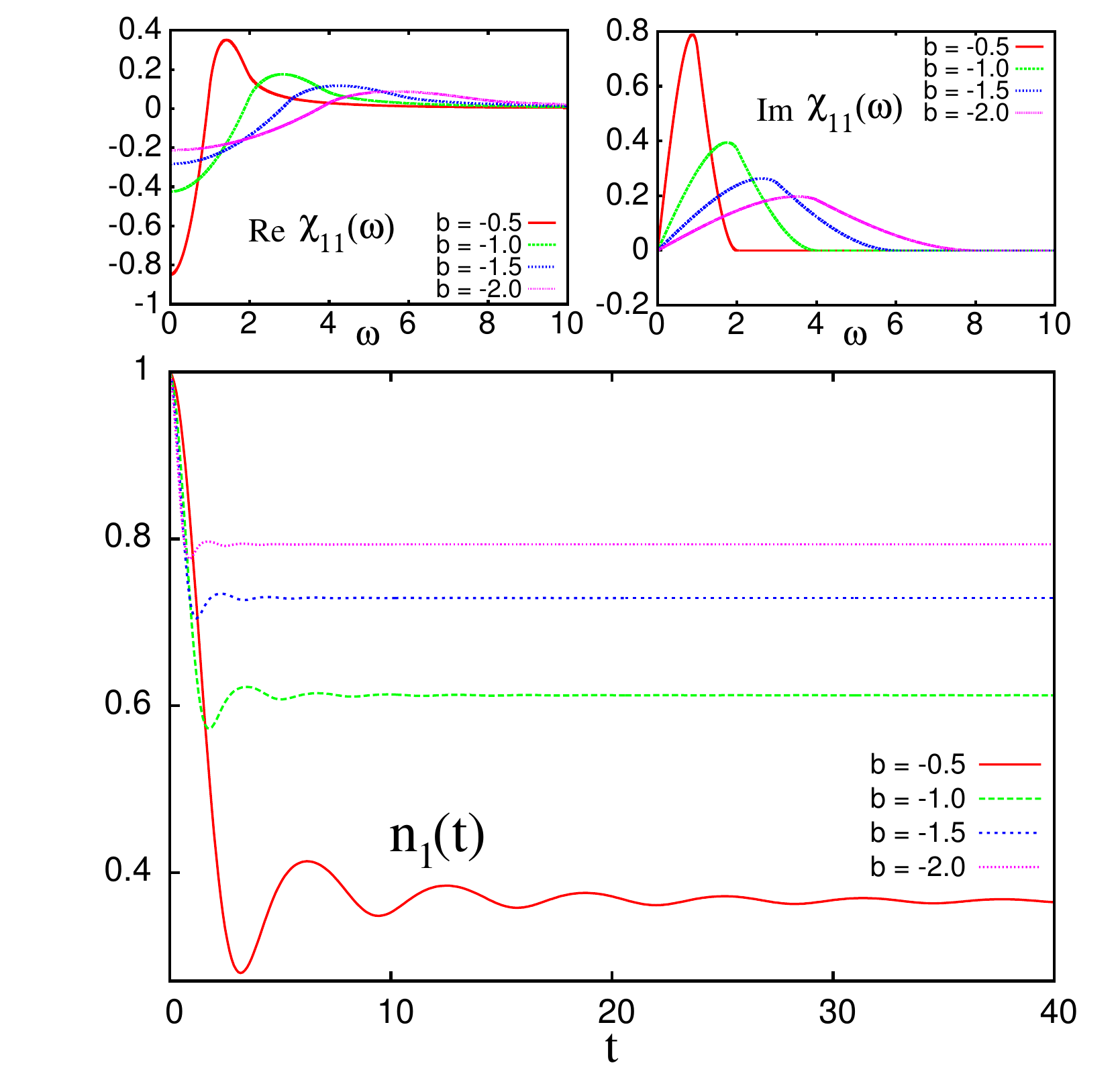}
\caption{Top left and right panels: 
The real and imaginary part  
of the dynamical response function. 
Bottom panel: The electron density at the terminal site of the 
TB chain as a function of time when the impurity potential $ U   = 0.5$ 
is suddenly switched on. The different curves correspond to different 
values of the hopping parameter $b = -0.5, -1.0, -1.5,-2.0$.}
\label{Fig2}
\end{figure}
The bottom panel of Fig. \ref{Fig2} shows how the image charge is 
built up in the lead. We plot the time-evolution of the density at 
the first site of a semi-infinite chain 
when the impurity-like potential $ U   = 0.5$ is suddenly switched 
on at time $t=0$ on site 1. The different curves correspond to different 
hopping parameter in the lead $b=-0.5,-1.0,-1.5,-2.0$. 
By increasing $b$ the frequency of the transient oscillations 
increases and the steady state is reached faster.
This behavior can be easily understood by inspecting the imaginary 
part of the density response  function $\chi_{11}(\w)$, top right panel  
of Fig.~\ref{Fig2}. In Appendix \ref{DRF} we show that this quantity has a maximum at 
$\omega \sim 2|b|$ which corresponds to the  oscillation   
frequency of the density. The width of the maximum grows like $2|b|$ 
and its inverse gives the {\em screening time}, i.e., the time-scale for 
the image charge formation.  Furthermore, from the top left panel we 
see that $\chi_{11}(\omega=0)$ behaves as $1/b$ 
 which is consistent with the larger induced charge in 
the long time limit.\\

\subsection{Many-body treatment}
\label{mb-treatment}

The ICM has not exact solution for the contacted case and to study it  
both in and out of equilibrium we use the non-equilibrium Green 
function (NEGF) method based on time-propagation  
of the embedded Kadanoff-Baym equations.\cite{Kadanoff,Myohanen.2008, 
Myohanen.2009, Stan.2009}
The basic quantity in the formalism is the one-particle Green 
function 
\be
G_{kl}(z,z') = -\mathrm{i} \frac{\Tr \left\{ 
\mathcal{T} [e^{-i\int_c \ud\bar{z} \ 
\hH(\bar{z})}\hat{c}_{k}(z)\hat{c}^{\dagger}_{l}(z')] \right\}}
{\Tr \left\{ e^{-i\int_c \ud\bar{z} \ \hH(\bar{z})}\right\}},
\ee
where we used the notation $\hat{c}_{k}$ and 
$\hat{c}^{\dagger}_{l}$ to denote electron annihilation and creation 
operators either in the molecule or in the leads.
In the above definition
$z,z'$ are the time indices on the Keldysh contour $c$, 
$\mathcal{T}$ is the time-ordering
operator on the Keldysh contour and $\Tr\{...\}$ signifies
the trace over the Fock space of all many-body states. The Green function $G$ is the solution 
of the integro-differential equation of motion on the Keldysh contour
\be
\bigl[ \mathrm{i}\partial_z - h(z)\bigr]G(z,z') =
\delta(z,z') + \int_c \! \ud\bar{z} \ \Sigma[G](z,\bar{z})G(\bar{z},z'),
\label{embkbe}
\ee
where $h(z)$ is the Hamiltonian in the one-particle Hilbert space, 
$\delta(z,z')$ is the contour delta function and $\Sigma[G]$ is the 
self-energy kernel containing all the information on the many-body 
and embedding effects \cite{Myohanen.2008,Myohanen.2009}. For the purpose of
a practical implementation of the Hamiltonian (\ref{Hamiltonian}), 
we divide the system into interacting ($C$) and noninteracting ($\alpha$) regions 
and write the single-particle part and the interaction part of $C$ as (see Eq. (\ref{Hamiltonian})) 
\begin{widetext}  
\be\label{KBHam1}
[h]_{ij}(t)= \left(\begin{array}{cccc}
-2U^l+W^l(t) & \lambda^l & \lambda^l & 0 \\
\lambda^l & \e_{H}-U^l-U^r & 0 & \lambda^r \\
\lambda^l & 0 & \e_{L}-U^l-U^r & \lambda^r \\
0 & \lambda^r & \lambda^r & -2U^r+W^r(t)
\end{array}
\right)\,\,\, ; \,\,\,
[v]_{ij}=\left(\begin{array}{cccc}
0 & U^l & U^l & 0 \\
U^l & U_0 & U_{HL} & U^r \\
U^l & U_{LH} & U_0 & U^r \\
0 & U^r & U^r & 0\\
\end{array}
\right).
\ee
\end{widetext}
Using this notation, the Hamiltonian (\ref{Hamiltonian}) transforms into
\beq\label{KBHam2}
\hat{H} &=& \sum_{ij\in C}\sum_{\sigma=\uparrow\downarrow}h_{ij}(t)\hat{c}_{i\sigma}^{\dagger}\hat{c}_{j\sigma} 
+ \frac{1}{2}\sum_{ij\in C}\sum_{\sigma=\uparrow\downarrow} v_{ij}\hat{c}_{i\sigma}^{\dagger}\hat{c}_{j\sigma}^{\dagger}
\hat{c}_{j\sigma}\hat{c}_{i\sigma}\nonumber\\
&+& \sum_{\alpha=l,r} \sum_{ij \in \a} 
\sum_{\sigma=\uparrow\downarrow}  
[ h_{ij}^{\alpha} + \delta_{ij}W^{\alpha}(t)]\hat{c}_{\alpha 
i\sigma}^{\dagger}\hat{c}_{\alpha j\sigma} - \mu\hat{N},\nonumber\\
\eeq
where $C$ contains the molecular levels and also the terminal sites of the leads 
subjected to the bias voltages $W^{\alpha}(t)$. 
Furthermore, $\alpha=L,R$ are the noninteracting parts of the left and right leads. We 
choose $U_0 = U_{HL} = U_{LH} = 1$, $\mu = 0$, $\epsilon_H=-2$ and $\epsilon_L=-1$ 
throughout the rest of this article.\\

We will solve Eq.(\ref{embkbe}) with a 
Hartree-Fock (HF), second-Born (2B) and GW many-body self-energy. 
The quality of the 2B and GW self-energy has recently been assessed 
in the Anderson model\cite{anna-maija} by comparing the 
time-dependent current and density against time-dependent DMRG 
results.\cite{feiguin} Good agreement was found in the 
parameter regime that we discuss below.

\begin{figure}[t]
\centering
\includegraphics[scale=0.7]{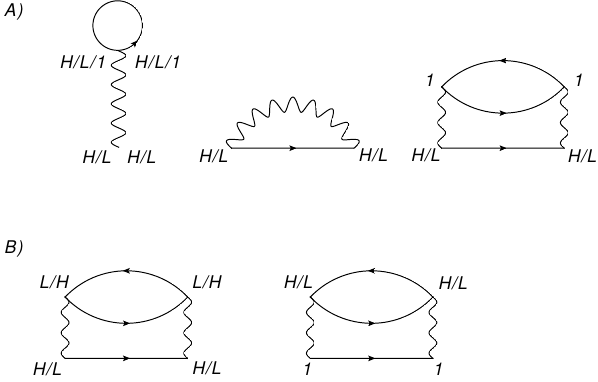}
\caption{Self-energy diagrams for $\lambda^{l}=\l^{r}=0$.
A) The only nonzero diagrams of the GW self-energy.
B) Second order self-energy diagrams that are zero. All higher order 
diagrams are also zero.}
\label{Fig3}
\end{figure}

It is instructive and useful for our later analysis to 
discuss the many-body approximations in the uncontacted case. 
Since the number of electrons in the $H$ and $L$ levels are conserved 
quantities in this case, the Green's function $G_{Hk}=\d_{Hk}G_{HH}$ and similarly 
$G_{Lk}=\d_{Lk}G_{LL}$ for all levels/sites $k$ of the system. 
The HF approximation consists of the first two diagrams in Fig. 
\ref{Fig3} A). The label of the vertices refer to the molecular 
levels $H$, $L$ and the terminal site of lead $\a$ (which is simply 
denoted by $1$ independently of $\a$). The 2B self-energy is obtained 
by adding to 
the HF self-energy the first bubble diagram and the second-order exchange diagram. 
For zero hybridization, however, the second-order exchange 
diagram vanishes since it either contains an off-diagonal 
element of the Green's function (which is zero) or a product of lesser and greater 
diagonal element of the molecular Green's functions, e.g., 
$G^{>}_{HH}G^{<}_{HH}$. Having the 
equilibrium system two electrons in $H$ and zero in 
$L$ it must be $G^{<}_{LL}=G^{>}_{HH}=0$. Consequently, only the first 
bubble diagram survives and the 2B self-energy diagrams are all 
displayed in Fig. \ref{Fig3} A). Note that if the external 
vertices of the bubble self-energy diagram lie on the terminal site of the leads 
[second diagram of Fig. \ref{Fig3} B)]
the diagram vanishes. This is a direct consequence of the fact that 
the polarization diagram with $H$ or $L$ as  vertices 
is proportional to the product $G^{>}_{HH}G^{<}_{HH}$ or $G^{>}_{LL}G^{<}_{LL}$ 
which is zero. For the same reason the first diagram of Fig. 
\ref{Fig3} B) is also zero. The physical origin of this result 
is that one cannot create particle-hole excitations on the molecules 
without changing the number of electrons in the $H/L$ levels.
The many-body self-energy in the GW approximation is $\S=iGW$ 
where the screened interaction $W$ is approximated as a geometric 
series of bare polarization diagrams connected by interaction lines.
Since the only bare polarization diagram is the particle-hole 
propagator going from 1 to 1,  the GW approximation coincides with the 2B 
approximation. In our ICM there is no direct 
interaction between two electrons on the terminal site of the leads.
We therefore expect  the 2B and GW approximations to perform similarly for 
small hybridizations.

\section{Competing time-scales and spectral properties}
\label{competing}

In the previous Section we have seen that there is a characteristic 
screening time to build up charge after the addition or removal of an electron to or from the 
molecule. 
In the case that the molecule is contacted to the leads there is another 
time-scale that plays a role.  This is the {\em relaxation time} 
to disperse the excess charge on the molecule into the leads.  
It is the ratio between the screening time and the relaxation time
that tells us how the system behaves under non-equilibrium conditions. 
The aim of this Section is to extract these time-scales from 
the equilibrium properties of the contacted system and to analyze the 
effects of the screening on the equilibrium spectral function.
This will help us to gain insight in the more complicated case of quantum 
transport discussed in the next Section.  
The analysis will be carried on using many-body Green's function methods 
since the contacted case is no longer analytically 
solvable.

\subsection{Screening and relaxation times}
\label{scr-rel}

\begin{figure}[t]
\centering
\includegraphics[scale=0.35]{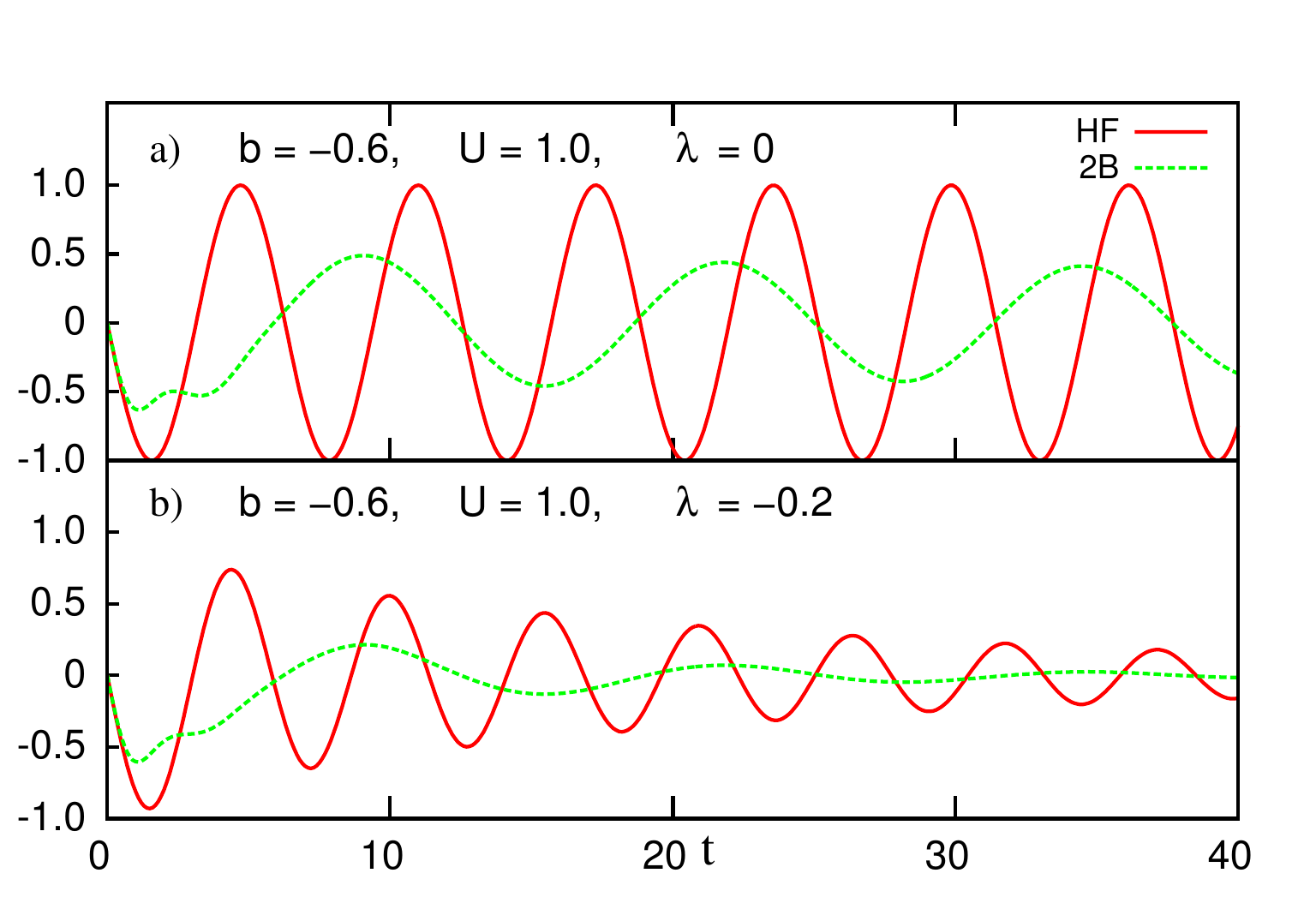}
\caption{Green function $G_{LL}^>(t,0)$ for a) HF and 2B with 
$b=-0.6$, $ U  =1$ and $\lambda=0$, b) HF and 2B with $b=-0.6$, $ U  =1.0$ and
$\l=-0.2$}
\label{Fig4}
\end{figure}

In this Section we study the response of the system to the sudden addition or removal of
an electron on the molecule within the HF, 2B and GW approximations. The response of an added
electron is encoded into the $G^>$ and $G^<$ Green functions 
which we can calculate within these many-body approximations both in real time and 
in frequency space. For instance, the LUMO Green's function 
$G_{LL}^>(t,0)=-i \langle \hat{c}_L (t) \hat{c}_L^\dagger (0) \rangle$
gives the probability amplitude of finding a particle on the  
LUMO level at time $t$ after being created at time $0$.  
In Fig.~\ref{Fig4} we plot the real part of this quantity.
This Green function oscillates with a characteristic frequency 
equal to the addition energy of an electron to the LUMO level. 
In the panel a) of Fig. ~\ref{Fig4} we compare the HF and 2B results for 
$\l^{l}=\l^{r}=0$, $U^{l}=0$, $U^{r}=U=1$ and $b=-0.6$. 
The correlated 2B curve exhibits a short transient with a 
characteristic time-scale $\sim 1/b$.
This transient has to be attributed to the build up of the image charge 
and its duration is consistent with the previous analysis of Fig. 
\ref{Fig2}. 
Note that no transient is visible in the HF approximation which 
therefore fails to describe the formation of the image charge.
\begin{figure}[t]
\centering
\includegraphics[scale=0.35]{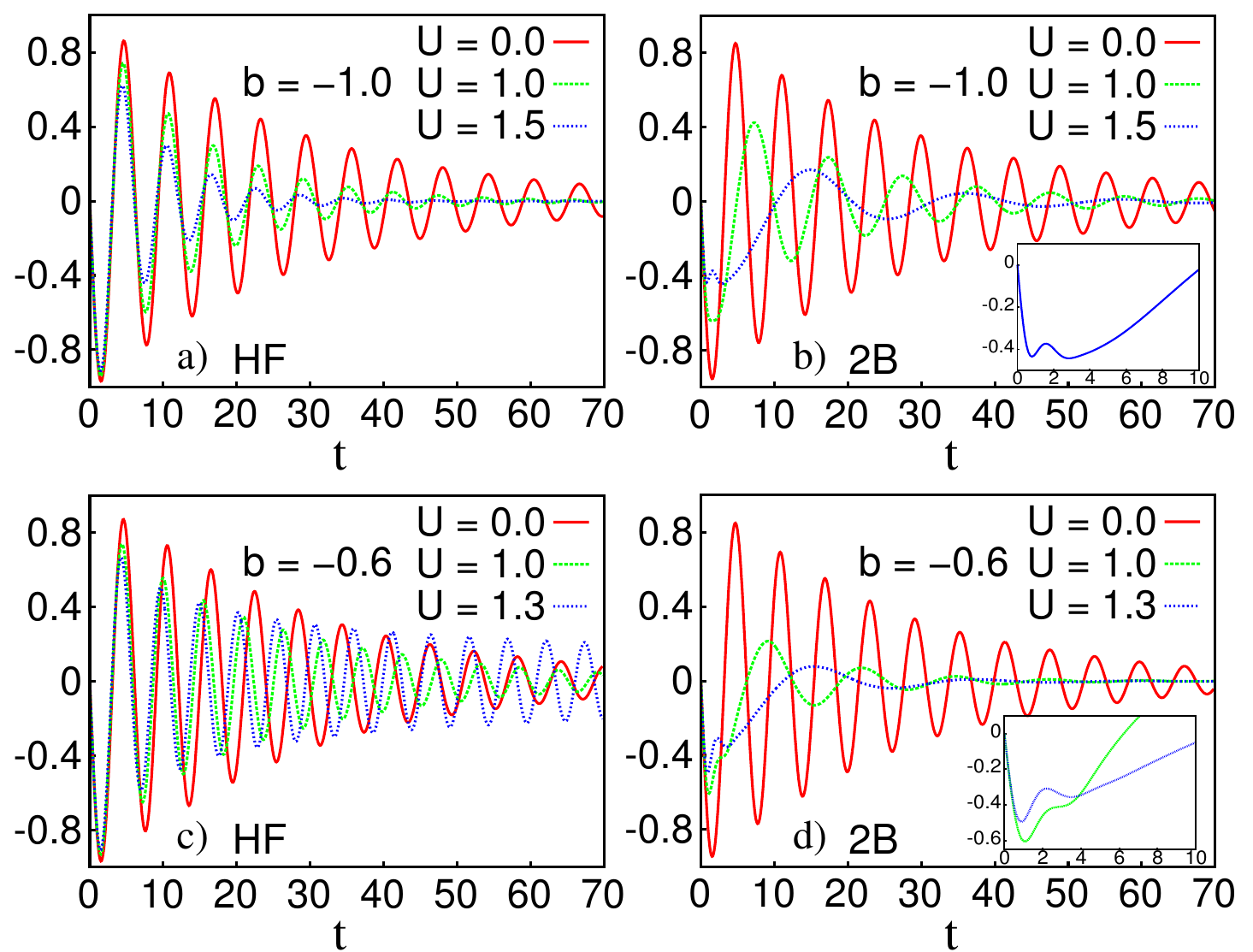}
\caption{Green function $G_{LL}^>(t,0)$ for HF and 2B with 
$b=-1.0$ (panels a) and b) ) and $b=-0.6$ (panels c) and d)) and $\lambda=-0.2$
and different values of $U$.}
\label{Fig5}
\end{figure}
The interplay between 
the screening time and the relaxation time can be investigated 
by contacting the molecule to the leads.
In the panel b) we consider the same parameters as in panel a) except 
for $\lambda^r=\l=-0.2$, i.e. the molecule is contacted to the right lead.
The main difference to the previous case is that both the
HF and 2B curves are damped (relaxation).
Similarly to the uncontacted case, there is no evidence of screening in the HF 
approximation.

Let us now address in more detail the dependence of the relaxation time
on the molecule-lead interaction $U$.
In most physical situations the 
band-width of the metallic leads is much larger than the  
molecule-lead coupling, $b \gg \lambda$. Then, 
for small values of $U$ 
the relaxation time is proportional to $\t_{\rm rel} \sim \Gamma^{-1} 
\sim |b|/\lambda^2$ . This time-scale depends
both on the molecule-lead coupling and the lead hopping. 
On the other hand, the 
screening time $\t_{\rm scr} \sim 1/ |b|$
is a property of the lead only and  
the ratio $\t_{\rm rel}/ \t_{\rm scr} \sim (b/\lambda)^2 $ is 
always larger than unity. If the interaction $U$ becomes comparable
to or larger than $b$ then this analysis is not valid anymore. In the 
Appendix \ref{renormalization} we show that
already at the HF level $U$ renormalizes the relaxation time according to 
$\t_{\rm rel} \sim \Gamma^{-1} (1-CU)^{2}$ where $C$ is a positive
real constant weakly dependent on $U$ for small values of $U$.
This is illustrated in the top panels of Fig.(\ref{Fig5})
where we display the real part of $G^>_{LL} (t,0)$ for $b=-1.0$ and 
$\lambda^r=\l=-0.2$ at HF and 2B level.
We see in both cases that by increasing the molecule-lead interaction $U$
we lower the relaxation time. The renormalization of the
lead coupling (or the embedding SE) also affects the positions of 
the molecular quasiparticle levels.
The renormalization leads to a small upward shift of the
LUMO level and a downward shift of the HOMO level, \emph{i.e.}, a slight opening
of the HOMO-LUMO gap. This is clearly visible in Fig. \ref{Fig5}a,
where we see a slight increase in the frequency of the LUMO oscillation 
when we increase $U$. In panel b) for 2B, on the other hand, we see a
much more drastic decrease of the oscillation frequency due to the image charge
effect which HF fails to describe properly.

We note that the small upward shift within the HF approximation of the LUMO level with increasing $U$ can lead to
an increase of the relaxation time when the level is close to the band edge.
This is because the upward shift pushes the LUMO level close to the band edge where
the imaginary part of the embedding self-energy decreases rapidly and compensates
the renormalization introduced by the interaction $U$.
In this case, the spectral peak describing the position and lifetime of the 
molecular quasiparticle level becomes also highly asymmetric and non-Lorenzian
which leads to a non-exponential decay of the Green function in real time. These features are
illustrated in Fig. \ref{Fig5}c where we consider the case
of lead hopping $b=-0.6$. The LUMO level for $U=0$ is located at $1$ which is close
to the band edge of $2|b|=1.2$. An increase of $U$ to 1.3 pushes the level very close
to the band edge and we then see a corresponding increase in relaxation time with
a non-exponential decay. In the case of 2B (panel d)) the image charge effect
pushes the level inwards, away from the band edge, and we see that the relaxation time 
again decreases with increasing $U$. Comparing panel d) to panel b) we see further
that increasing the lead hopping $b$ leads to a slight decrease of the image-charge 
effect (frequency change for $U \neq 0$) and increase of the relaxation time,
in agreement with the analysis of Section \ref{exsol} and the relation
$\t_{\rm rel} \sim \Gamma^{-1} \sim |b|/\lambda^2$.

\begin{figure}[t]
\centering
\includegraphics[trim = 10mm 00mm 00mm 0mm, clip, scale=0.37]{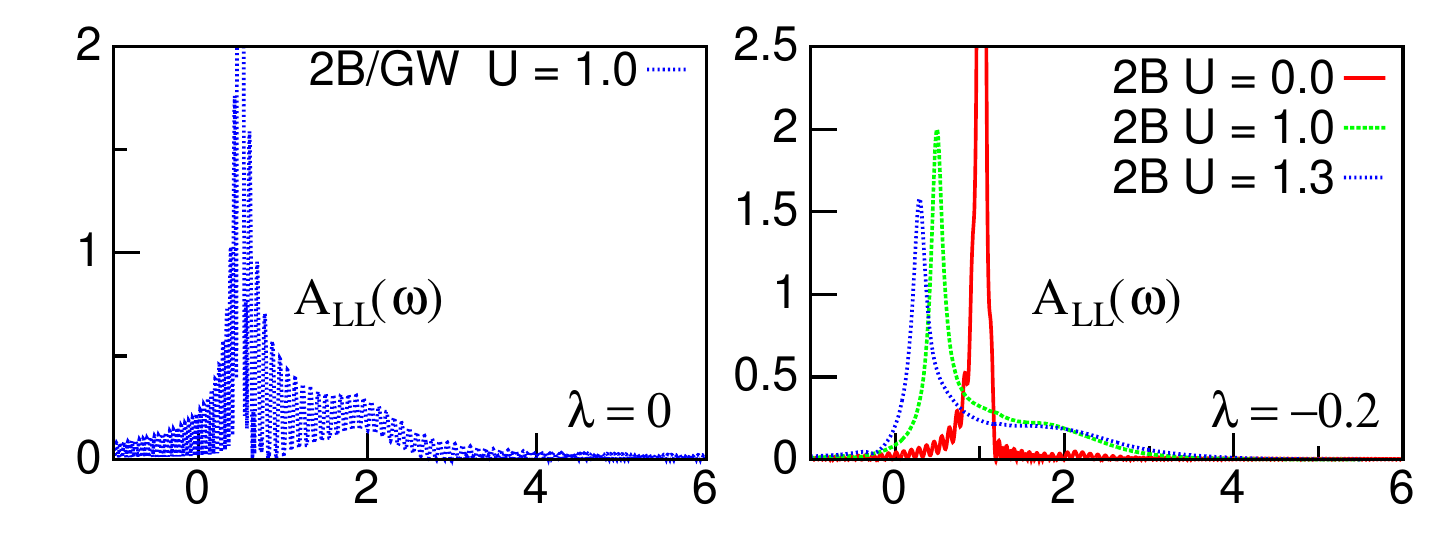}
\caption{LUMO spectral function in the 2B approximation for 
$b=-0.6$, $U=1.0$ and $\lambda=0$ (left panel), $b=-0.6$, $\lambda=-0.2$ and $U=0,1.0,1.3$ (right 
panel).}
\label{Fig6}
\end{figure}

The difference between the HF and the correlated results in real time 
translates into a different spectral structure in frequency space. In  
Fig.~\ref{Fig6} we display the quasiparticle spectral functions (see Eq. \ref{matsf})
for the LUMO level, $A_{LL}(\w)$, corresponding to the Green functions $G_{LL}^>(t,0)$.
This is done for the 2B approximation using $b=-0.6$ and 
different values of $U$.
For the uncontacted case the 2B result coincides with the GW
result, as discussed in Section 
\ref{mb-treatment}. The corresponding spectral function for $U=1.0$ is displayed
in the left panel while in the right panel we have $\l=-0.2$ and we plot 
the spectral function for different values of $U$.
The very fast oscillations in the left panel are due to the finite time interval in 
the Fourier transform. They are not present in the right panel due to damping of 
the Green function in the contacted case.
Besides the main peak 
located at the electron affinity we observe a shoulder of  
width $\Delta \approx 4|b|$ at higher energies. 
At finite hybridization $\lambda=-0.2$ this shoulder is smoother and 
partially merges with the main peak. The 
shoulder originates from the particle-hole
continuum of excitations  induced by
the sudden addition of an electron to 
the LUMO state. They are these excitations which allow for the  
dynamical screening of the extra charge on the 
molecule. In mathematical terms the shoulder arises by Fourier 
transforming the initial transient of the 2B curve in Fig. 
\ref{Fig4} and \ref{Fig5}. Since no transient was observed in HF, the HF 
spectral function will consist only of a main peak at the electron 
affinity. Both 2B and GW incorporate the correct physics through the 
polarization diagram of Fig. \ref{Fig3}A, which nicely illustrate 
how an extra electron on the LUMO can excite a particle-hole on the 
terminal site of the leads. For small hybridizations  
the polarization is approximatively equal to the response 
function of Fig. \ref{Fig2}, which explains the width  $4 |b|$ of
the shoulder. We further see in the right panel of Fig. \ref{Fig6} that while the peak
moves leftward with increasing $U$ the width of the plateau remains 
roughly constant at $4 |b|$. The screening time is therefore independent of the
molecule-lead interaction.

The message to take home is that the molecule-lead interaction have  
a large impact on the ratio $\t_{\rm rel}/ \t_{\rm scr}$ and, in principle, 
can turn it to be smaller than one. This kind of exotic situations 
would occur in leads with flat bands as, e.g., those modeled by 
Tasaki.\cite{tasaki} In most metallic systems this is not the 
case and in the remainder of this paper we will study the regime  
$\t_{\rm rel}/ \t_{\rm scr}>1$.

\subsection{Equilibrium spectral function}
\label{eqsf}

\begin{figure}[t]
\centering
\includegraphics[scale=0.67]{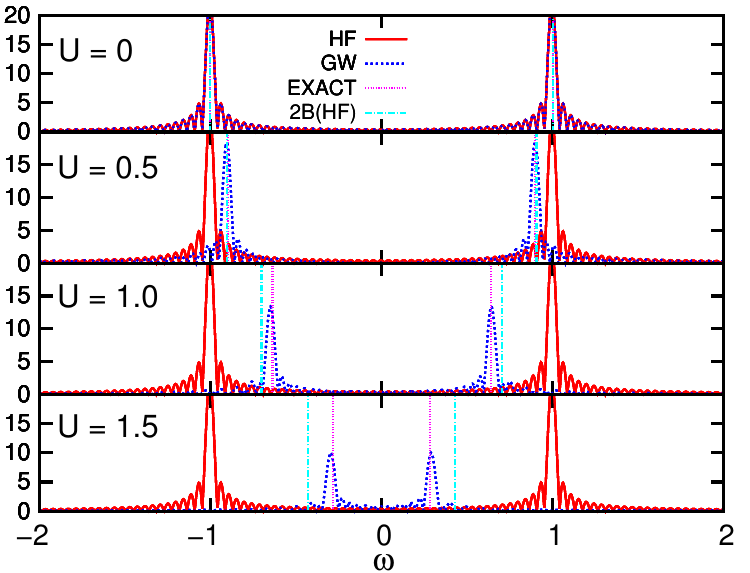}
\caption{HF and GW  equilibrium  
spectral functions $A_{\textnormal{mol}}(\omega)$ for $\lambda=0$, $b = -1.0$ and $ U  
=0, 0.5, 1.0, 1.5$.  
The vertical lines indicate the exact and 2B(HF) peak positions.}
\label{Fig7}
\end{figure}

\begin{figure}[t]
\centering
\includegraphics[scale=0.67]{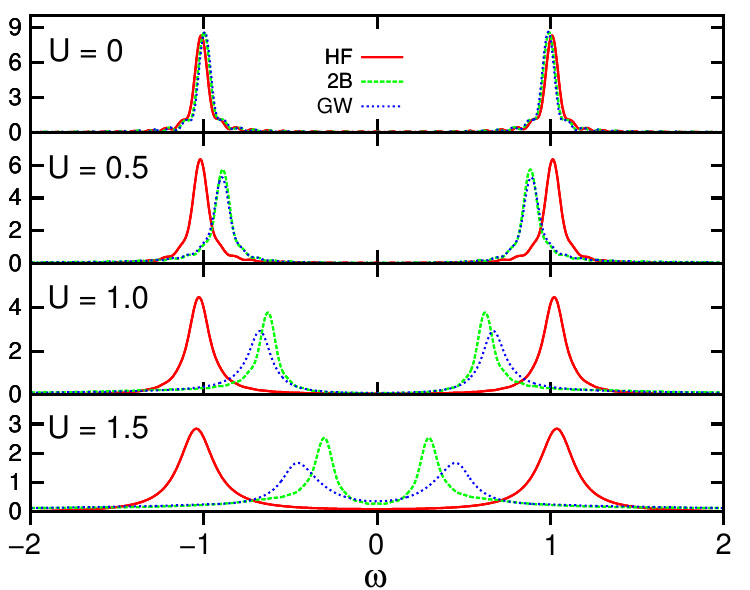}
\caption{HF, 2B  and GW 
equilibrium spectral functions $A_{\textnormal{mol}}(\omega)$ for $\lambda=-0.2$, $b = -1.0$ and $ U  
=0, 0.5, 1.0, 1.5$.}
\label{Fig8}
\end{figure}

In this Section we investigate the effects of screening on the spectral features 
of the molecule in equilibrium. 
We calculate the molecular spectral function $A_ {\textnormal{mol}}(\w)$ as a sum of the 
projected spectral components $A_ {ii}(\w)$ as
\be
A_ {\textnormal{mol}}(\w)=-\frac{1}{\p}\sum_{i=H,L}\,\Im [G_{ii}^\mathrm{R}(\w)],
\label{matsf}
\ee
for $U^{l}=\l^{l}=0$ and for zero and finite hybridization 
$\l^{r}=\l$ with the right lead. The results are shown in 
Fig. \ref{Fig7} for $\l=0$ and in Fig. 
\ref{Fig8} for $\l=-0.2$.

Let us start by analyzing the performance of the HF approximation. 
The first observation is that for $\lambda=0$ the HOMO-LUMO gap and 
the intensities of the   
peaks remain unchanged as the interaction strength $U^{r}=U$ increases.
This can easily be understood from the explicit form of the HF 
HOMO and LUMO energies 
\beq
\label{HFsolutions}
\epsilon_H^{\mathrm{HF}} &=& (\epsilon_H- U  ) + n_HU_0 + 2n_LU_{HL} 
+ 2n_{1} U  ,\\
\epsilon_L^{\mathrm{HF}} &=& (\epsilon_L\,- U  ) + n_LU_0 + 
2n_HU_{HL} + 2n_{1} U  .
\eeq
At half-filling the average density $n_{1}=n_{1r}$ of the right terminal 
site is $1/2$ and hence the dependence on $U$ cancels off. 
In the case of finite hybridization $\lambda=-0.2$ (Fig. 
\ref{Fig8}), the HF peaks shift slightly outwards and broaden 
due to the renormalization of the embedding self-energy 
(or equivalently, the renormalization of the hybridization $\l\ra\l+UG^{<}_{H1}$, see Appendix \ref{renormalization}).
It is then clear that for $\lambda = 0$ the intensities do not change since $G_{H1}=0$.
Similar renormalization effects has been observed in Ref. \onlinecite{Sade.2005}. 
In the HF approximation the self-energy of the $H/L$ levels couples only 
to the density at the terminal site of the lead and thus misses entirely 
the particle-hole coupling responsible for the screening.

The situation is radically different in the correlated 2B and GW 
approximations. In both cases the HOMO-LUMO gap, corresponding to the 
difference $A-I$ between the electron affinity and the ionization 
potential, narrows in agreement 
with the discussion of Section \ref{exsol}. The added/removed electron and its image charge  
bind together, thereby decreasing/increasing the addition/removal 
energy. The stronger is the interaction $ U  $ and the larger is the gap 
reduction. In the 2B and GW approximations the added/removed electron 
couples not only 
to the density but also to the particle-hole 
continuum of the lead. It is through this latter coupling that the charged system
can lower its energy by exciting particles from occupied to 
unoccupied levels of the charge-neutral system. The resulting 
effect  is to accumulate or deplete charge in the neighborhood of the 
terminal site, i.e., to screen the excess charge of the molecule. 
Note also that in the case $\lambda \neq 0$, the coupling to the particle-hole continuum provides 
an extra channel for quasi-particle scattering and induces quasiparticle
broadening to the spectral peaks. The differences between the 
uncontacted and contacted spectral functions must be attributed to  
charge transfer processes and the consequent formation of image 
charges in the molecule.  This molecular polarization effect was 
recently found to reduce the HOMO-LUMO gap even further.\cite{kr1.2009}

\begin{figure}[t]
\centering
\includegraphics[scale=0.34]{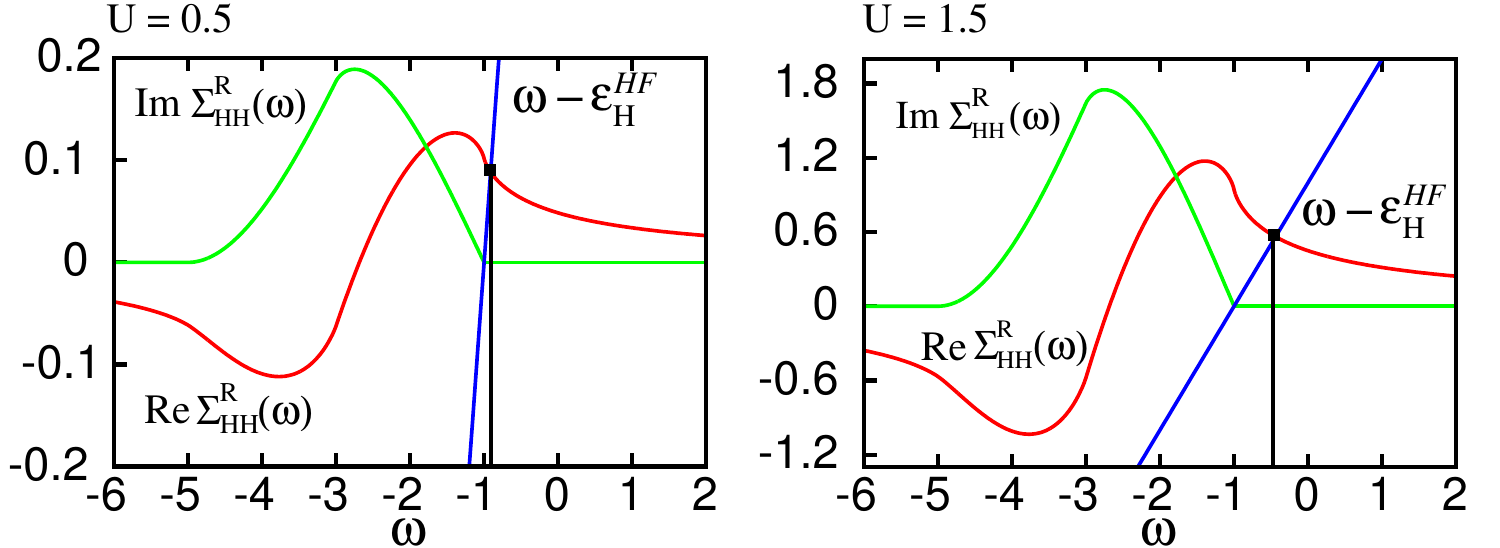}
\caption{The real and imaginary components of the 2B(HF) self-energy for 
the molecular HOMO level with $ U  =0.5$ (left panel) and $ U  
=1.5$ (right panel). The rest of the parameters are $U_{HL}=U_0=-b=1$ 
and $\l=0$.}  
\label{Fig9}
\end{figure}

To assess the quality of the correlated approximations and the 
importance of self-consistency we display in 
Fig.~\ref{Fig7} the position of the 
exact $H/L$ peak (calculated from the Hellman-Feynman theorem) 
as well as the position of the peaks as obtained from a one-shot 2B calculation 
with HF Green function ( denoted with 2B(HF) ). 
As can be seen from Fig.~\ref{Fig7} 
the GW results are in very good agreement with the exact ones. 
The position of the spectral peaks in the 
correlated approximations are obtained from the quasiparticle equation 
\be
\label{QPE}
\omega - \epsilon_i^{\mathrm{HF}} - 
\textnormal{Re}\left\{\Sigma_{ii}^{\mathrm{R}}(\omega)\right\} = 0, 
\ee
where $\Sigma_{ii}^{\mathrm{R}}(\omega)$ is the retarded many-body 
self-energy projected onto the $i=H,L$ molecular level. 
In Fig.~\ref{Fig9} we  display the graphical solution of 
Eq. (\ref{QPE}) with  
2B(HF) self-energy and $i=H$. For this plot we have chosen  
$U_{HL}=U_0=-b=1$ and $ U  =0.5$ (left panel) and $ U  =1.5$ (right 
panel). Already one iteration of the 
self-consistency cycle captures the correct trend. The zero of Eq. 
(\ref{QPE}) moves toward higher energies with increasing $U$. An 
analogous calculation for the LUMO level shows that the zero moves 
toward lower energy. In conclusion, the inclusion of polarization 
effects into the self-energy has two main effects in equilibrium: the redistribution of 
the spectral weight due to particle-hole excitations (satellite spectrum) and the 
collapse of the HOMO-LUMO gap. As we shall see, the situation is 
radically different out of equilibrium.

\section{Quantum transport: Short-time dynamics}
\label{sh-time-dyn}

In order to investigate the short-time transport properties of the 
system of Fig. ~\ref{Fig1}
we  consider  $\l^{l}=\l^{r}=\l$ and $U^{l}=U^{r}=U$. 
We will analyze the transient 
dynamics after the sudden switch-on of a bias $W^{l}=-W^{r}=W$ in the leads. 
Note from Eqs. (\ref{KBHam1}) and (\ref{KBHam2}) that the bias is applied also to the 
terminal (interacting) sites of the leads. We will refer to the left/right 
current as the current flowing through the left/right interacting-noninteracting 
interfaces correspondingly. In all simulations 
we set $\l=-0.2$ and hence work in the weak tunneling regime to 
highlight correlation effects. 

\subsection{HF approximation}
\label{HF-short-time}

\begin{figure}[t]
\centering
\includegraphics[scale=0.35]{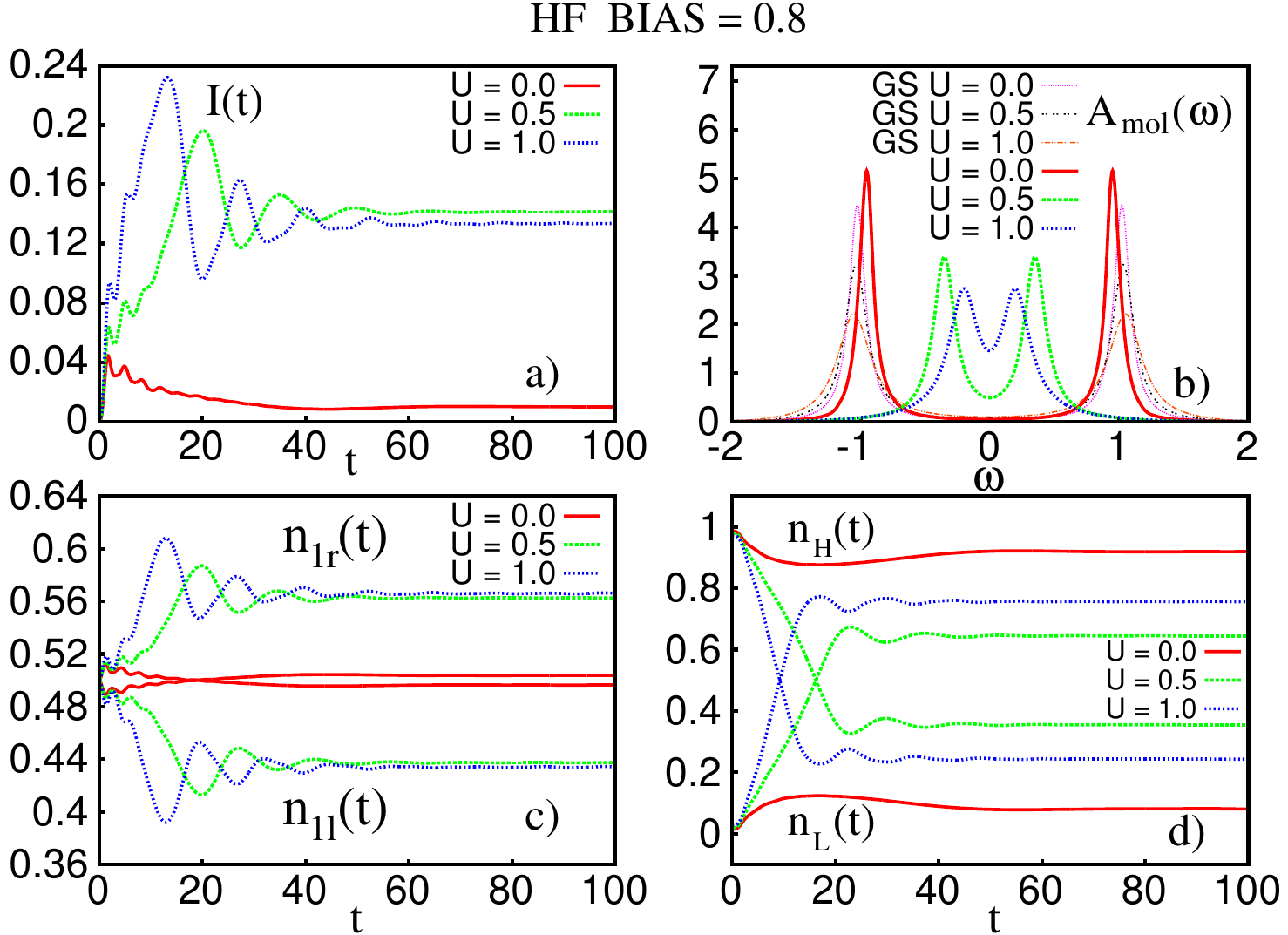}
\caption{a) Time-dependent right current for  $ U   = 0.0, 0.5,1.0$.  
b) Ground state (GS) and steady state spectral function of Eq. (\ref{matsf}).
c) Time-dependent densities $n_{1r}(t)$ and $n_{1l}(t)$ at the 
terminal sites. 
d) Time-dependent HOMO and LUMO densities $n_H(t)$ and $n_L(t)$. 
In all the plots the simulations have been performed within the HF 
approximation with bias $W^l = -W^r = 0.8$.}
\label{Fig10}
\end{figure}

\begin{figure}[t]
\centering
\includegraphics[scale=0.35]{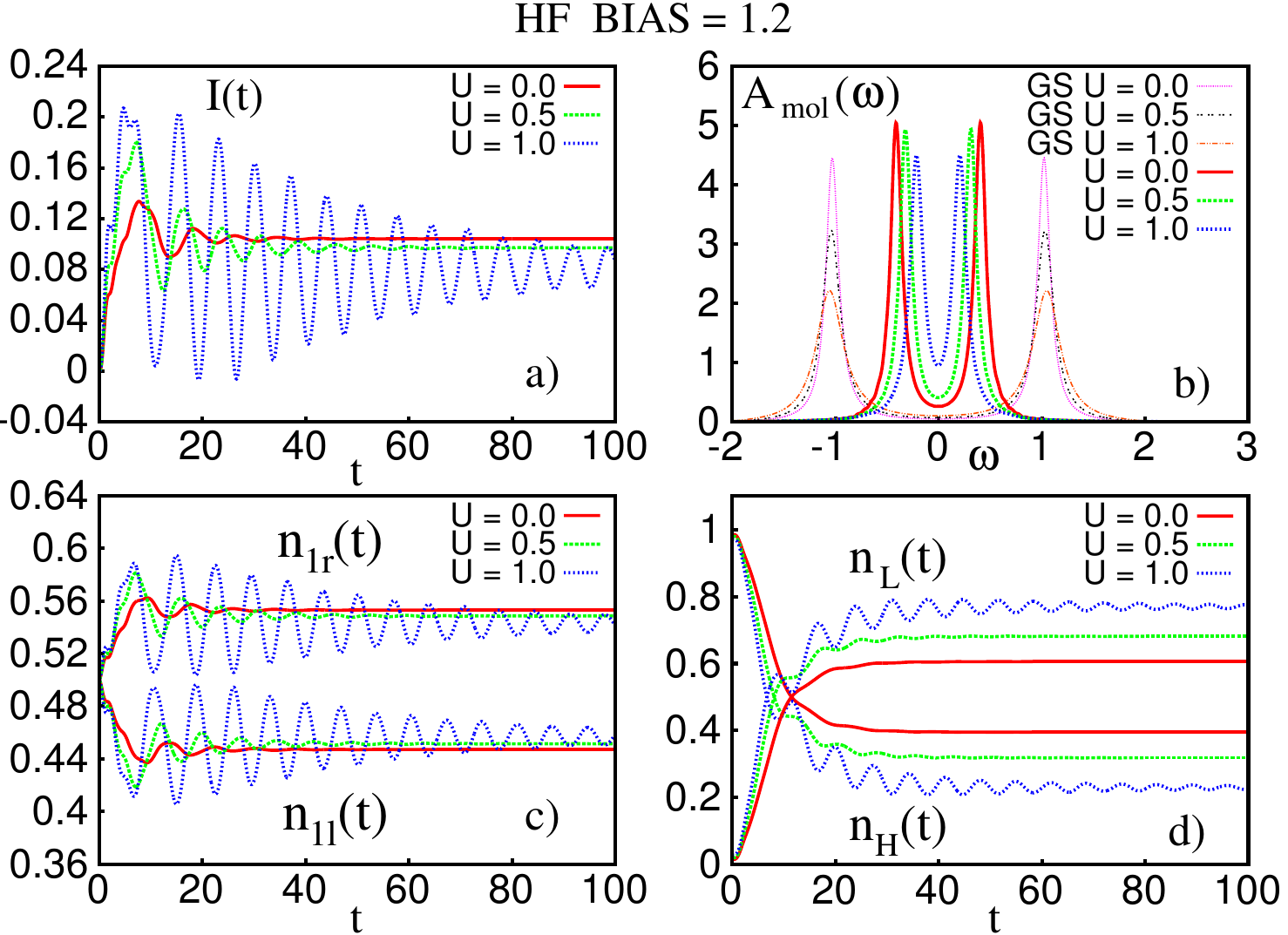}
\caption{a) Time-dependent right current for  $ U   = 0.0, 0.5,1.0$.  
b) Ground state (GS) and steady state spectral function of Eq. (\ref{matsf}).
c) Time-dependent densities $n_{1r}(t)$ and $n_{1l}(t)$ at the 
terminal sites. 
d) Time-dependent HOMO and LUMO densities $n_H(t)$ and $n_L(t)$. 
In all the plots the simulations have been performed within the HF 
approximation with bias $W^l = -W^r = 1.2$.}
\label{Fig11}
\end{figure}

In Figs. \ref{Fig10} and \ref{Fig11} we show the 
time-dependent currents (panel a), ground state and nonequilibrium steady  
state spectral functions (panel b), terminal site 
densities (panel c) and HOMO/LUMO densities (panel d) for the HF 
approximation with molecule-lead interaction $ U   = 0.0, 0.5, 1.0$.  
In Fig.~\ref{Fig10} we consider the ``small'' bias case 
$W^{l} = -W^{r} = 0.8$ for which the equilibrium $H/L$ levels 
$\e_{H/L}^{\textnormal{HF}}=\mp 1$ remain 
outside the bias window while in 
Fig.~\ref{Fig11} the bias is set to 
$W^{l} = -W^{r} = 1.2$ so that the equilibrium $H/L$ levels lie inside 
the bias window. 

For zero molecule-lead interaction, $U=0$,  and small bias  
the current flowing through the system is almost 
zero, see Fig.  \ref{Fig10}a, in agreement with the fact that 
the $H/L$ levels are outside the bias window. A finite current instead 
sets in for large bias, see Fig.  \ref{Fig11}a. The physics is 
here very similar to that of the non-interacting resonant transport regime.
On the other hand, the current increases substantially at finite $U$  
for  small  bias.
Furthermore, increasing $ U  $ the frequency and the amplitude of the 
oscillations in the current and density becomes larger.  
We recall  that in the HF approximation the equilibrium 
quantities are fairly independent of $U$. These results show that at finite bias
the situation is completely different. 

To understand the differences between the equilibrium and the 
non-equilibrium case we observe that the gap  in the non-equilibrium 
spectral function {\em reduces} considerably at finite  $ U  $. 
For instance  Fig.~\ref{Fig10}b shows that for $ U  
=0.5$ and $ U  =1.0$  
the $H/L$ levels have already entered  the bias 
window  $[-0.8,0.8]$. 
To trace back the physical origin of this effect we 
write the HF energies of the $H/L$ levels 
\beq
\epsilon_H^{\mathrm{HF}} &=& \epsilon_H-2 U   + U_0n_H + 2U_{HL}n_L + 
2 U  [n_{1r} + n_{1l}],\nonumber\\
\epsilon_L^{\mathrm{HF}} &=& \epsilon_L-2 U   + U_0n_L + 2U_{HL}n_H + 
2 U  [n_{1r} + n_{1l}],\nonumber
\eeq 
where we took into account that the molecule is now connected to both 
leads. The terms containing an explicit dependence on $U$ cancel 
off since  
the sum of the terminal site densities, $n_{1r}(t) + n_{1l}(t)$, remains 
roughly at its ground state value during the entire time evolution, 
see panels c). Thus, it is not the lead polarization which affects 
the level positions but rather the polarization of the molecular region, 
i.e., the difference $n_{H}-n_{L}$. 
The panels d)  indicate that the molecular polarization increases as $ U  $ 
becomes large.  This analysis shows that in the HF approximation the 
reduction of the gap induced 
by  $ U  $ has the same nature  observed earlier\cite{Myohanen.2008} 
and  has nothing to do with the image charge effect. However, as we will see later,
this effect already has a big impact on the resulting current-voltage characteristics.

\begin{figure}[t]
\centering
\includegraphics[scale=0.35]{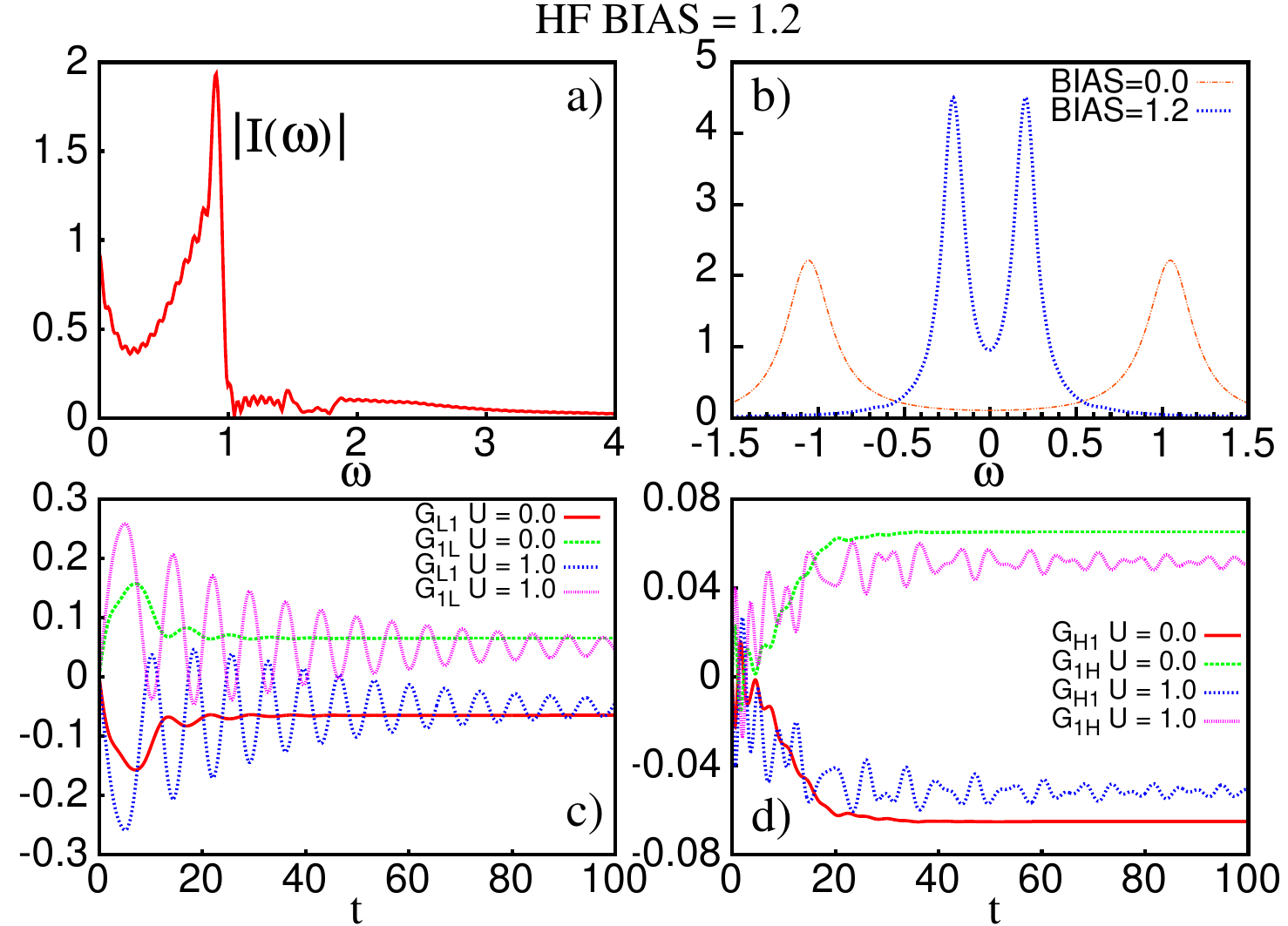}
\caption{a) Fourier transform of the transient 
current of Fig. \ref{Fig11} with $ U   = 1.0$.  
b) Ground state and steady-state spectral function 
$A_{\textnormal{mol}}(\w)$ for $ U   = 1.0$.
c) Time-dependent density matrix components $G_{L,1l}(t,t^+)$ and 
$G_{1l,L}(t,t^+)$ for $ U   = 0.0$ and $ U   = 1.0$ 
d) HF time-dependent density matrix components $G_{H,1l}(t,t^+)$ and 
$G_{1l,H}(t,t^+)$ for $ U   = 0.0$ and $ U   = 1.0$. 
In all the plots the simulations have been performed within the HF 
approximation with bias $W^l = -W^r = 1.2$.}
\label{Fig12}
\end{figure}
The main frequency of the oscillations in the transient density and 
current originate from the  
electronic transitions from the left electrochemical potential 
$\m^{l}=\e_{F}+W^{l}$ to 
the LUMO level and also from the HOMO level to the
right electrochemical potential $\m^{r}=\e_{F}+W^{r}$ (for the symmetric 
bias considered here these transitions have the same energy). This can easily be verified by 
calculating the discrete Fourier transform  
of the transient current, $I(\omega)$. In Fig.~\ref{Fig12} we 
show $I(\omega)$ for $ U  =1.0$ and the large bias case 
$W^{l}=-W^{r}=1.2$ (panel a) along with the ground state and 
nonequilibrium   
steady-state spectral function (panel b). 
The Fourier transform $I(\w)$ exhibits a sharp peak at
$\omega\approx1.0$ with a smearing towards lower frequencies   
down to $\omega\approx 0.2$. The smearing is a direct consequence of
including the transient part of $I(t)$ in the Fourier transform. 
The value of $\epsilon_{H/L}^{\mathrm{HF}}$ is 
$ \mp 1.0$ in equilibrium while it is about $\mp 0.2$ at the steady state, see 
Fig. \ref{Fig12}b. 
As the HOMO-LUMO gap collapses, the 
transition energy between the left/right electrochemical potential 
and the LUMO/HOMO level changes   
from $0.2$ to $ 1.0$. The aforementioned smearing towards low 
frequency is the fingerprint of the dynamical renormalization of the 
transition frequency.
Another consequence of the collapse of the steady-state gap with 
increasing $U$ is that  
$\epsilon_{H/L}^{\mathrm{HF}}$ moves further away from $\m^{l/r}$ 
where the density of states has a 
square-root divergence (resonance condition). This is clearly 
illustrated in Fig.~\ref{Fig11}b. The further away the levels are from 
resonance the harder it is for electrons to tunnel, which in turn 
implies a larger oscillation amplitude  and a smaller average current. 

The transient oscillations are also visible in the 
off-diagonal components  
of the time-dependent density matrix, $G_{ij}(t)\equiv 
G_{ij}(t,t^+)$,  which is displayed in Figs.~\ref{Fig12}c 
and~\ref{Fig12}d for $ U   = 0$ and $ U   = 1.0$.  
The component $G_{L,1l}(t)$ and $G_{1l,L}(t)$ oscillate with 
the  same  main frequency as the current and densities. The same 
holds true for  $G_{H,1r}(t)$ and $G_{1r,H}(t)$ (not shown). 
On the contrary  $G_{H,1l}(t)$ and $G_{1l,H}(t)$ 
have a very weak high frequency 
component superimposed to the main frequency. 
Initially the HOMO level is fully occupied and electronic 
transitions from $\m^{l}$ to  $\epsilon_{H}^{\mathrm{HF}}$ are 
blocked. Similarly, the LUMO level is initially empty, so there are no 
electronic  transitions from $\m^{r}$ to $\epsilon_{L}^{\mathrm{HF}}$.
As the time passes, however, the HOMO occupation decreases while the 
LUMO occupation increases and these transitions become possible.  
They are located around $1.4$ and $2.0$  
and can be seen in the Fourier transform
of the current (the current is indeed given in terms of 
off-diagonal elements of the density matrix).
Even though present, the transitions between the HOMO level and the LUMO 
level are extremely small since there is no direct hopping between 
the two levels. 
\begin{figure}[t]
\centering
\includegraphics[scale=0.65]{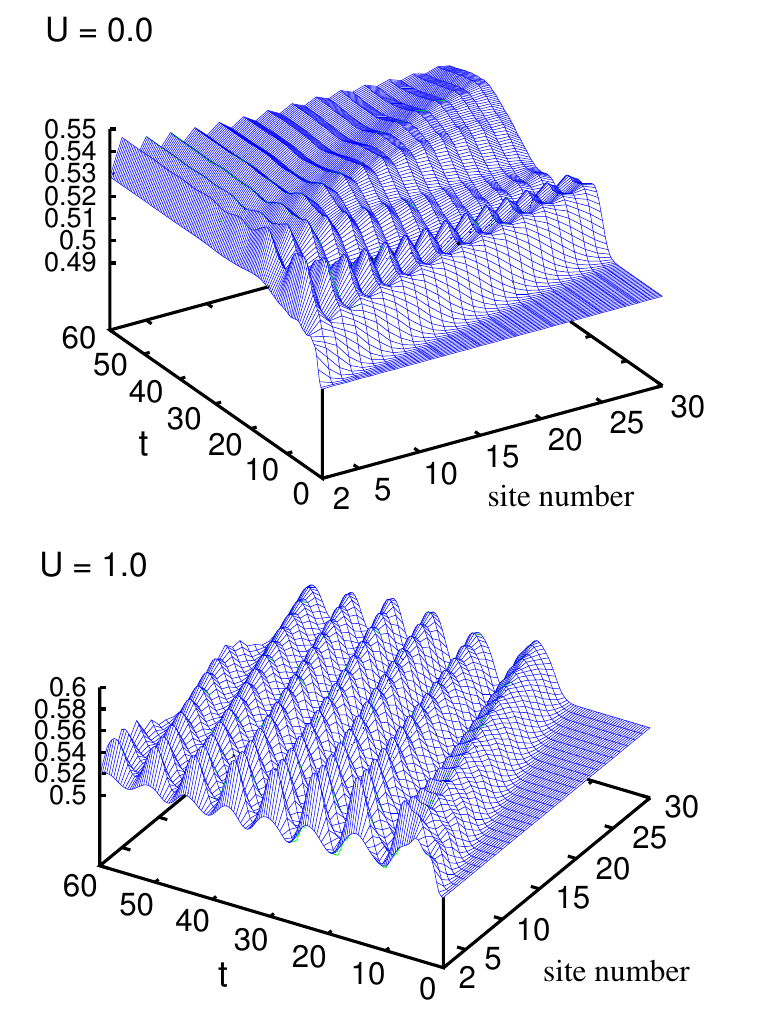}
\caption{Time-dependent density in the non-interacting part of the 
right lead for $ U  =0.0$ (top panel) and $ U  =1.0$ (bottom panel). 
The simulations have been performed within the HF 
approximation with bias $W^l = -W^r = 1.2$. Site number $2$ corresponds 
to the first noninteracting site in the right lead.}
\label{Fig13}
\end{figure}

The sudden switch-on of the bias gives rise to density shock waves in 
the leads with features similar to the density at the terminal sites.
In Fig.~\ref{Fig13}  
we show the transient dynamics of the HF density in the noninteracting part
of the right lead for $ U  =0.0$ (top panel) and for $ U  
=1.0$ (bottom panel) when the bias voltage is $W^{l}=-W^{r} = 1.2$. 
The shock wave reaches site $j$ after a time $j/v_{F}$ where in our 
case the Fermi velocity $v_{F}=2b$. No matter how far site $j$ is the 
density at this site exhibits damped oscillations whose initial 
amplitude and relaxation time is 
independent of $j$ and increases with $U$. 

\subsection{Correlated approximations}

\begin{figure}[t]
\centering
\includegraphics[scale=0.35]{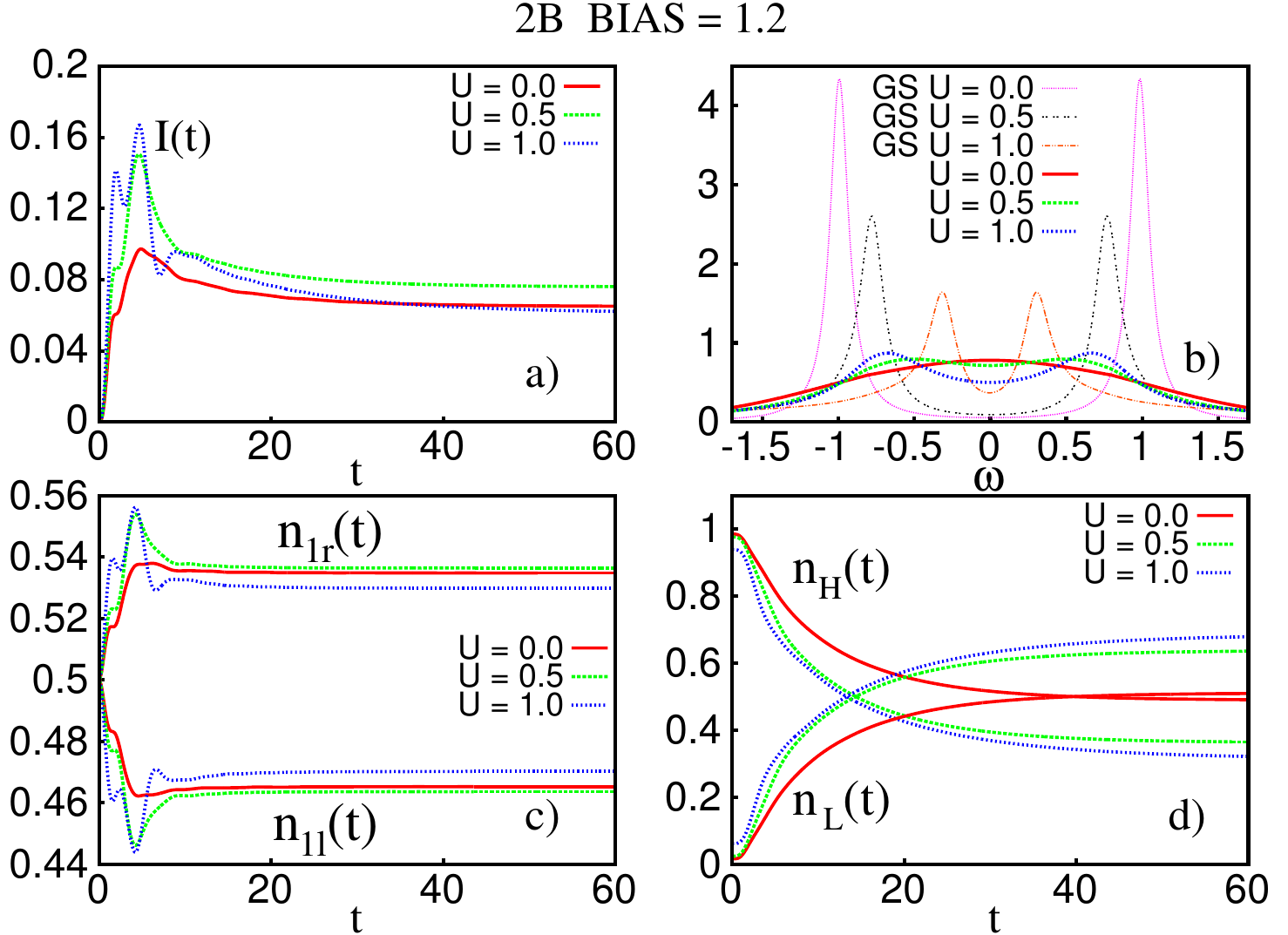}
\caption{a) Time-dependent right current for  $ U   = 0.0, 0.5,1.0$.  
b) Ground state (GS) and steady state spectral function of Eq. (\ref{matsf}).
c) Time-dependent densities $n_{1r}(t)$ and $n_{1l}(t)$ at the 
terminal sites. 
d) Time-dependent HOMO and LUMO densities $n_H(t)$ and $n_L(t)$. 
In all the plots the simulations have been performed within the 2B 
approximation with bias $W^l = -W^r = 1.2$.}
\label{Fig14}
\end{figure}
\begin{figure}[t]
\centering
\includegraphics[scale=0.35]{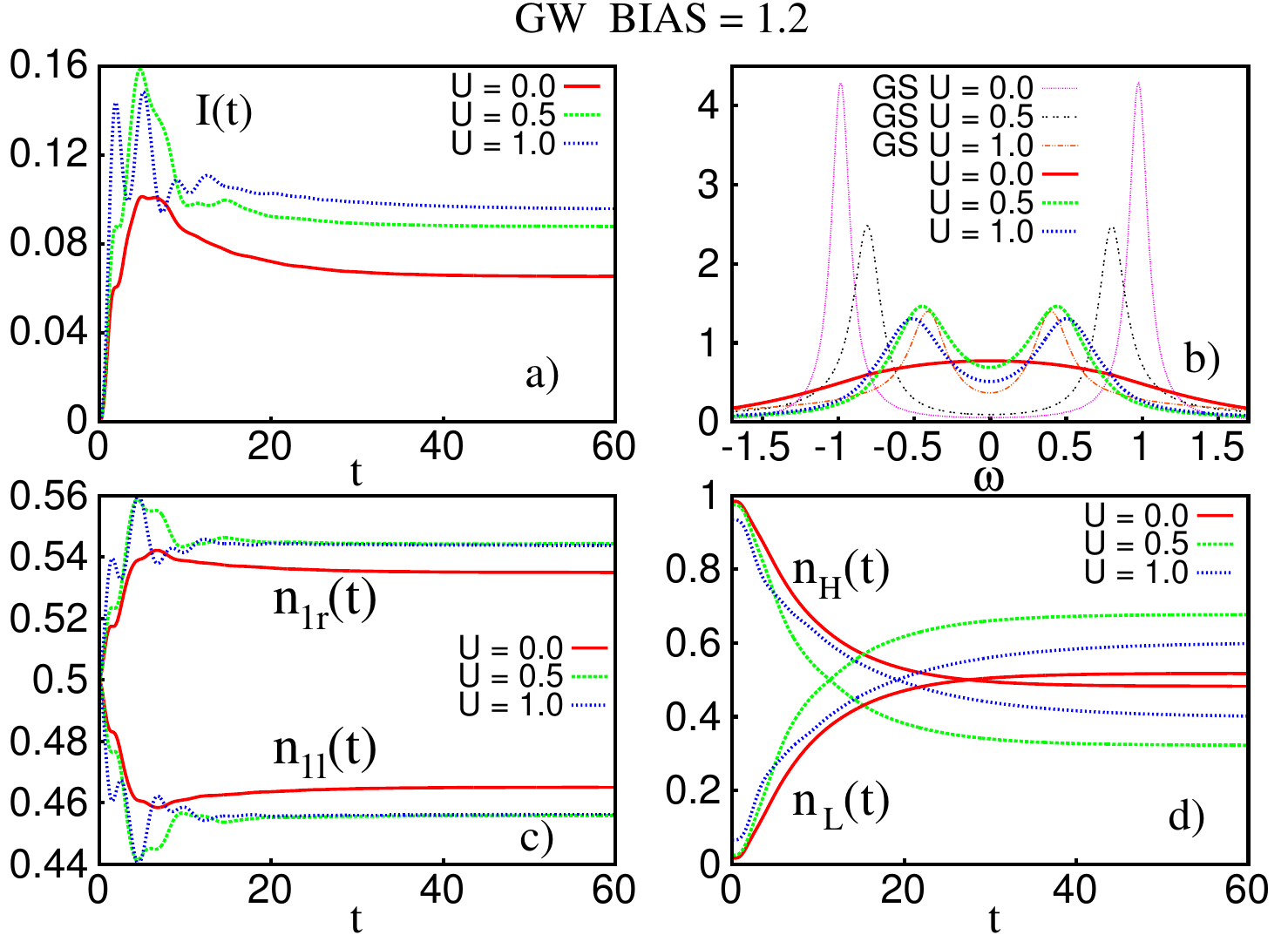}
\caption{a) Time-dependent right current for  $ U   = 0.0, 0.5,1.0$.  
b) Ground state (GS) and steady state spectral function of Eq. (\ref{matsf}).
c) Time-dependent densities $n_{1r}(t)$ and $n_{1l}(t)$ at the 
terminal sites. 
d) Time-dependent HOMO and LUMO densities $n_H(t)$ and $n_L(t)$. 
In all the plots the simulations have been performed within the GW 
approximation with bias $W^l = -W^r = 1.2$.} 
\label{Fig15}
\end{figure}

The inclusion of correlations changes considerably the physical 
picture. Let us focus on the large bias case  $W^{l}=-W^{r} = 1.2$ 
and calculate the same quantities as in Fig. \ref{Fig11} but 
within the 2B and GW approximation. The results are displayed in 
Fig. \ref{Fig14} and ~\ref{Fig15} respectively. The first 
important feature is that the relaxation time is much shorter than in 
the HF case due to
the many-body broadening of the HOMO and LUMO 
levels, see panels b).
In the same panels we also show the ground state (GS) spectral function
for the same values of $ U  $. As expected the GS gap between the 
HOMO and LUMO peaks reduces with increasing $U$ due to the image charge effect.
In the biased system for $U=0$ the bias dependent gap closing \cite{kr3.2008,Myohanen.2008}
brings the levels so close to each other that we can observe only one very broad peak.
Interestingly and 
surprisingly, the effect of increasing $U$ in the biased system is to 
{\em open} the gap and to sharpen the spectral peaks.
In the 2B approximation with molecule-lead interaction 
$ U   = 1.0$ the nonequilibrium steady-state gap is even 
larger  than the ground-state gap. The GW approximation 
attenuates the gap opening compared to the 2B approximation, 
but the sharpening of the peaks is well 
visible also in this case. The gap opening in the out-of-equilibrium 
system has never been reported before and, as we shall see below, 
has profound consequences on the $I-V$ curve.

\section{Quantum transport: long time dynamics}
\label{long-tim-dyn}

In this Section we investigate the effects of the image charge on the 
long-time dynamics of the  lead-molecule-lead system  within the HF, 2B 
and GW approximation. As we shall see a non-trivial post-transient 
dynamics develops at the HF level. The inclusion of 
correlations does always bring the system in a steady-state regime. 
We will show how this regime is attained and calculate current and 
densities in the steady state for different bias voltages and 
molecule-lead interaction.

\subsection{HF approximation and post-transient dynamics}

\begin{figure}[t]
\centering
\includegraphics[scale=0.7]{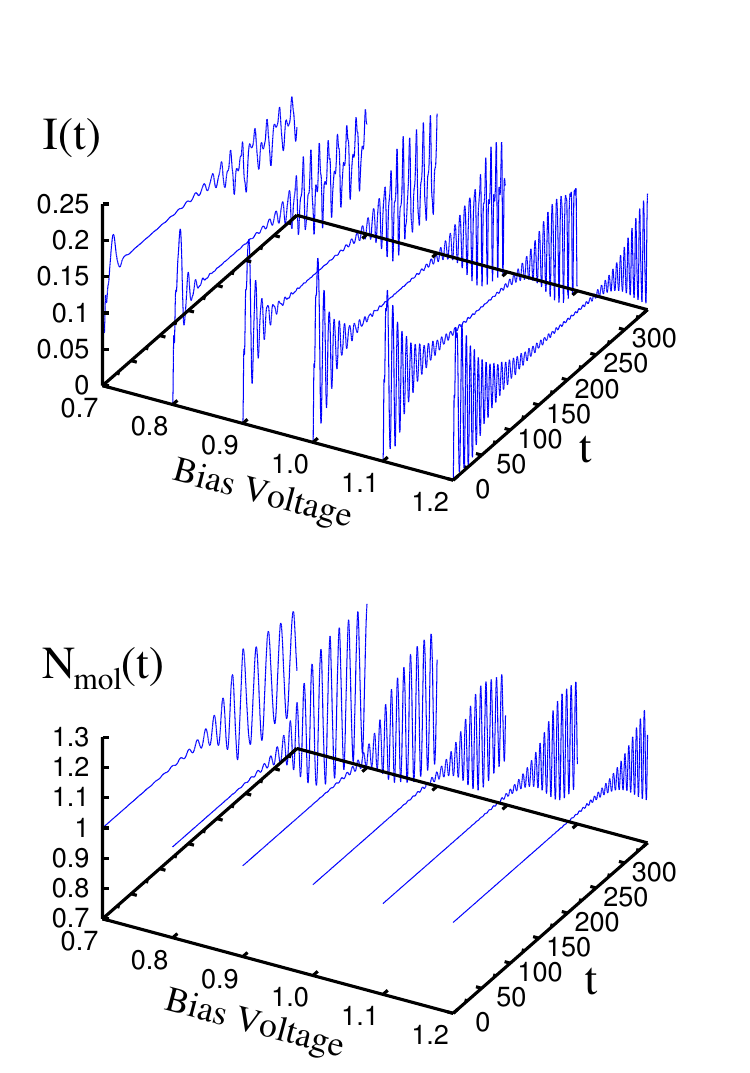}
\caption{Time dependent right current (top panel) and total number of 
particles in the molecule (bottom panel) for bias voltages in the 
range $[0.7,1.2]$ and molecule-lead interaction $ U  =1.0$.}  
\label{Fig16}
\end{figure}

We focus on the large bias regime and strong  
molecule-lead interaction $ U  =1.0$. In the previous Section we 
showed that current and densities seem to relax after the transient 
behavior induced by the sudden switch-on of a bias voltage. 
However, extending further the propagation time-window something unexpected 
occurs. We find that the steady state is metastable and oscillations 
with increasing amplitude develop to then stabilize in a  
periodic state.
In  Fig.~\ref{Fig16} we display long-time simulations of the 
right current $I(t)$ (top panel) and   
the total number of particles in the molecule $N_{\rm mol}(t)$ (bottom panel) for 
bias voltages in the range $[0.7,1.2]$.
In this range the equilibrium HOMO and LUMO levels lie in the bias window.
The frequency of the oscillations increases  
as the bias voltage is increased, which is a clear indication that 
the dominant transitions are those between the leads and the molecular levels. 

\begin{figure}[t]
\centering
\includegraphics[scale=0.36]{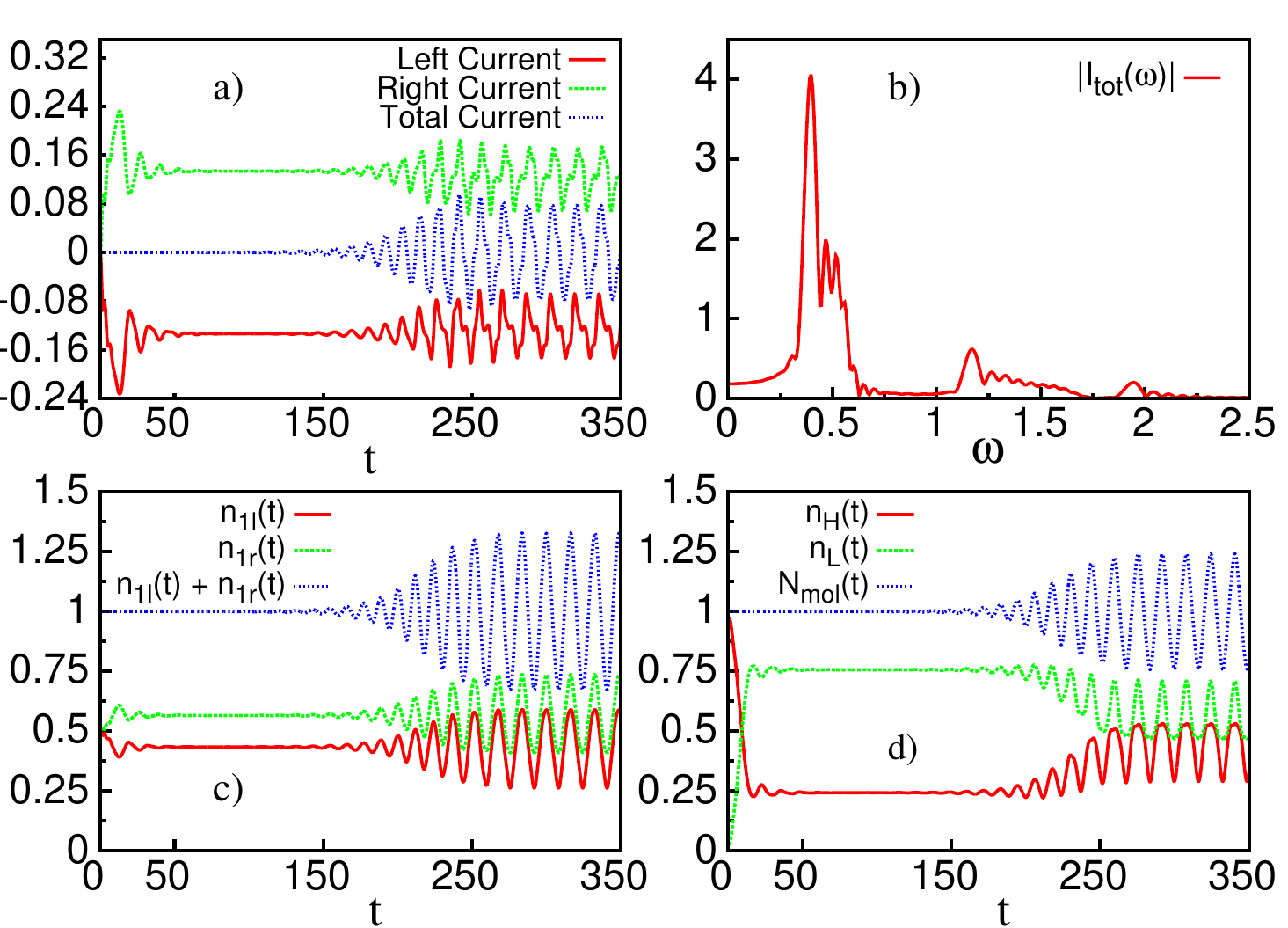}
\caption{a) Time-dependent left, right and total current. b) Fourier 
transform of the total current. c) Time-dependent terminal site  
densities. d) Time-dependent 
HOMO and LUMO densities and the total number of particles in the 
molecule. In all  
panels $ U  =1.0$ and the bias voltage is $W^{l}=-W^{r}=0.8$.} 
\label{Fig17}
\end{figure}

In  Fig.~\ref{Fig17} we display the time-dependent 
left and right currents (panel a) as well as the terminal-site 
densities (panels c) and molecular densities (panel d) for 
$W^{l}=-W^{r}=0.8$. According to these results the 
post-transient periodic state corresponds to a sequence of charge 
blockades with opposite sign of the electron-liquid acceleration 
(time-derivative of the current) between two 
consecutive blockades. The oscillations are therefore due to a charge 
sloshing between the molecular levels and the terminal sites. 
The metastability of a steady-state solution in which current 
and densities are given by the average value of the time-dependent 
results is due to the combination of the constant flow of electrons 
from left to right and the self-consistent nature of the Hartree-Fock 
potential. 
Finally we emphasized that the amplitude of the ac current superimposed to the dc 
current depends on where the current is measured.

In Fig. \ref{Fig17}b we report the Fourier transfrom  of the 
total current, $I_{\rm tot}(\omega)$. The main peak at $\omega\approx0.4$ 
is smeared out toward higher frequencies up to $\omega\approx 
0.6$, indicating  the occurrence of
electronic transitions between levels whose position changes 
dynamically in time. We 
also observe higher frequency satellites   
arond $\omega\approx 1.2,1.8$. These satellites occur 
exactly at   
the positions of odd harmonics of the main frequency. 
The absence of even harmonics is due to the fact 
that the external driving field is 
an odd function in space.

The persistent oscillatory behaviour reported in this Section is 
most likely an artifact of the HF 
approximation and, as we shall see in the next Section, disappears in 
the 2B and GW approximations.  Within HF the system knows 
only the instantaneous density and there is no damping mechanism  to
wash out  the oscillations. These oscillations are sustained by the 
finite bias voltage and originate from   the 
instantaneous Coulombic feedback.

\subsection{Steady-state properties: HF, 2B and GW approximation}

\begin{figure}[t]
\centering
\includegraphics[scale=0.37]{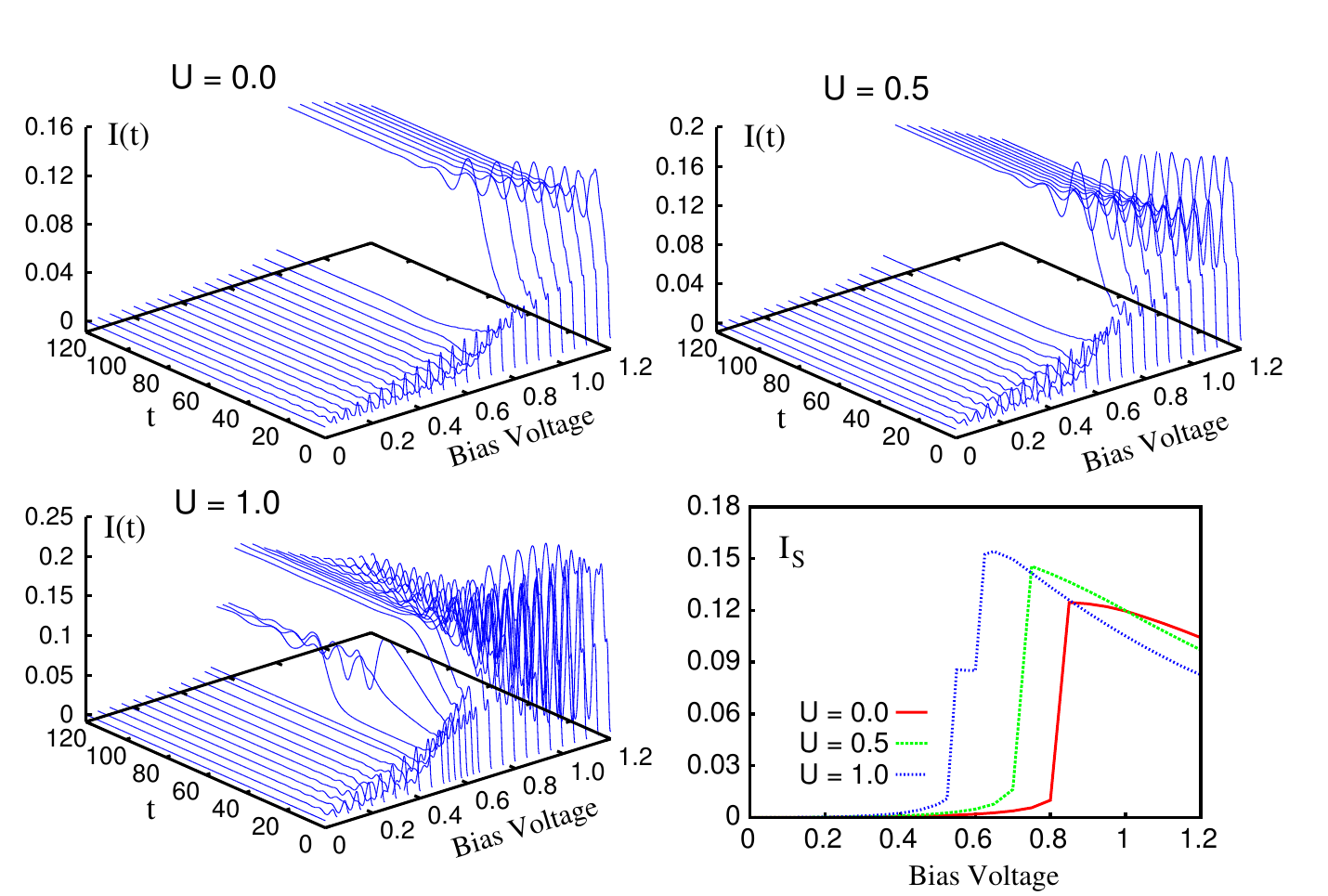}
\caption{HF time-dependent right current for different interaction  
strengths $ U   = 0, 0.5$ (top left and right panels) and $ U  =1.0$ 
(bottom left panel). The $I-V$ curves extracted from the long-time 
limit are displayed in the bottom right panel.} 
\label{Fig18}
\end{figure}

In Fig. ~\ref{Fig18} we show the HF time-dependent currents  
and the resulting $I-V$ characteristic
(bottom right panel)  for  
different interaction strengths $ U   = 0.0, 0.5, 1.0$. 
Since the HF currents for $ U   = 1.0$ do not attain a steady 
state for large enough bias, the $I-V$ characteristic is in 
this case calculated with the dc part of the current 
(average value).
The inclusion of the molecule-lead interaction 
deforms the $I-V$ characteristics 
dramatically. 
Firstly, increasing the interaction strength, 
the  threshold is shifted towards smaller bias values.
Secondly, increasing the interaction strength up to 
$ U  =1.0$ gives rise to  an extra step in the $I-V$ curve.
The shift of the $I-V$  step towards smaller biases is related 
to the gap closing mechanism which in the HF approximation is 
entirely due to the intramolecular interactions $U_{0}$ and 
$U_{HL}$, see Section \ref{HF-short-time}.\\

\begin{figure}[t]
\centering
\includegraphics[scale=0.55]{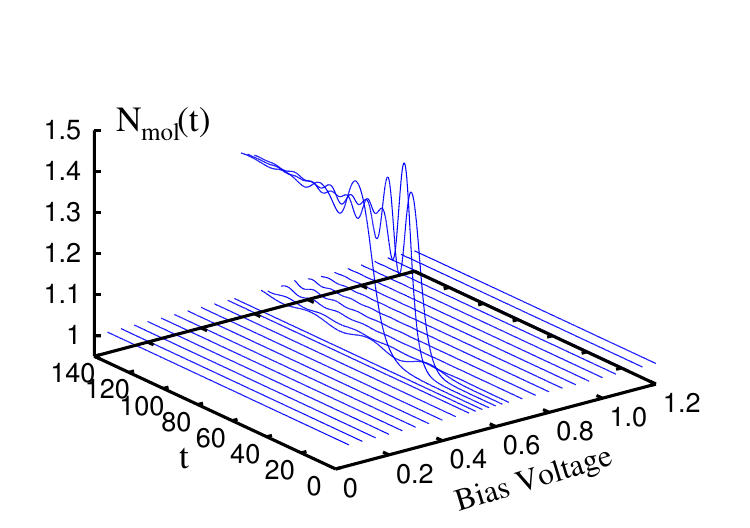}
\caption{Time-dependent number of electrons on the molecule, 
$\mol(t)$, versus the applied bias voltage.  
HF approximation with $ U  =1.0$.}
\label{Fig19}
\end{figure}

The extra step in the HF $I-V$ curve (bottom panel of 
Fig.~\ref{Fig18}) corresponds to a charged state of the 
molecule. In Fig. \ref{Fig19} we plot 
the number of particles (per spin) in the molecule, $\mol$, for 
interaction $ U   = 1.0$. There exists a narrow 
window of applied biases $W^{l}=-W^{r} \in [0.55,0.6]$ for which 
 $\mol\approx 1.35$. We have also checked (not shown here) that this window 
can be extended by increasing the molecule-lead coupling $\lambda$.
The excess molecular charge 
produces a Hartree barrier on the terminal sites  
which prevents the current to increase, see plateau in the  $I-V$ 
curve for $U=1$. 
As the bias  becomes larger electrons gain enough energy to overcome the barrier 
and the current increases again.   

\begin{figure}[t]
\centering
\includegraphics[scale=0.35]{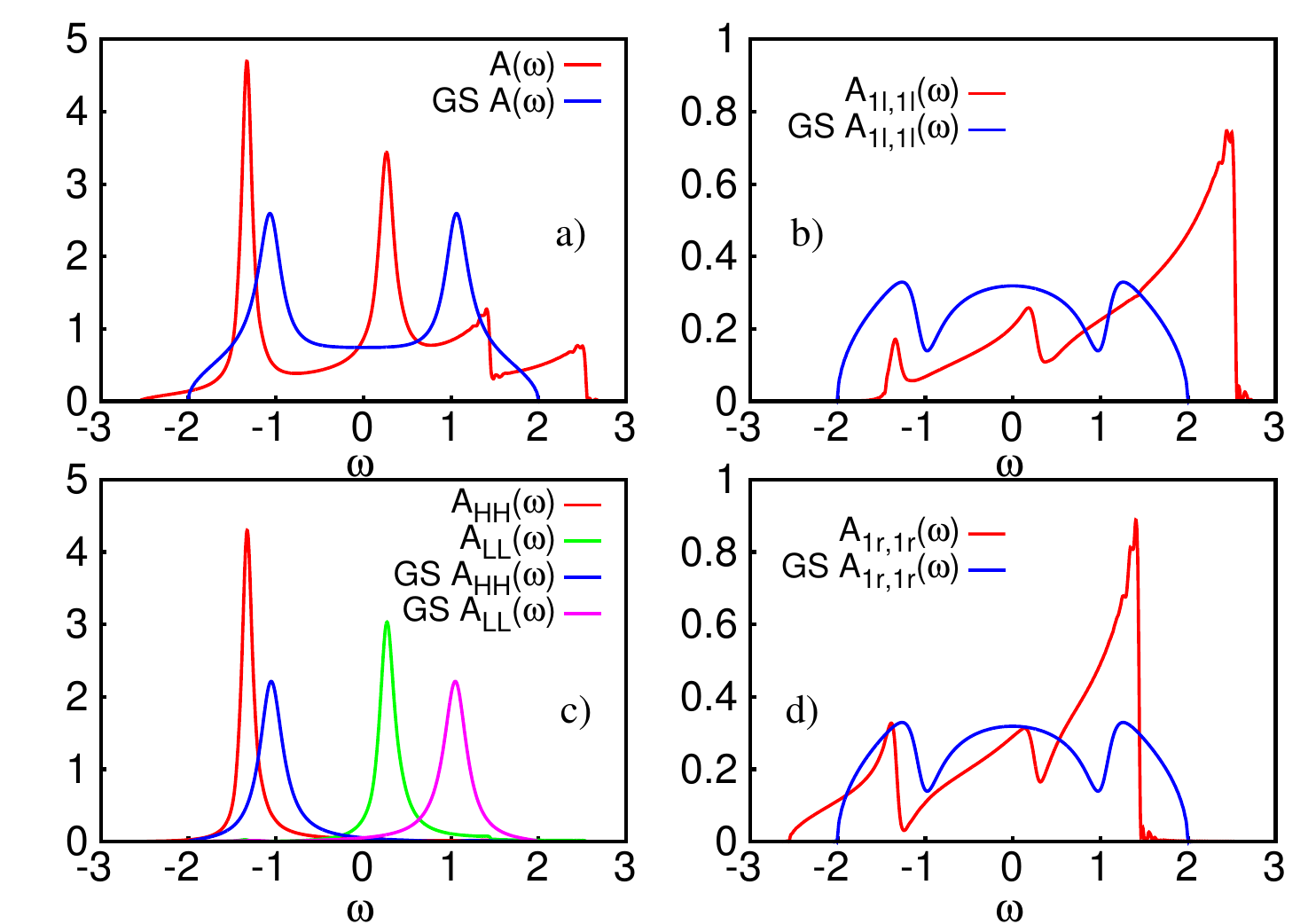}
\caption{Ground state (GS) and steady-state spectral functions in the 
HF approximation for  $ U  =1.0$ and bias $W^{l} = -W^{r} = 0.55$.
a) Full spectral function of the interacting region. 
b) Spectral function on the terminal site of the  
left lead. c) HOMO and LUMO spectral functions. 
d) Spectral function on the terminal site of the  
right lead.} 
\label{Fig20}
\end{figure}

The excess charge on the molecule  changes also the  spectral 
function. 
In Fig.~\ref{Fig20} we  
plot the full spectral function of the interacting region as well 
as the local spectral 
functions of the HOMO, LUMO and the terminal sites in the ground 
and steady state for $ U = 1.0$ and for bias $W^{l} = 
-W^{r} = 0.55$ within the HF approximation.
The HF spectral function of the charged 
molecule exhibits two sharp structures close to the left and right 
band edges (they are separated by $W^{l}-W^{r} = 1.1$). 
The induced Hartree barrier pushes  
electrons away from the terminal sites and gives rise to well 
localized hole states. The structure of the peaks is indeed similar 
to that of a split-off state (anti-bound state) which forms 
in the presence of an external positive potential at the endsite of a 
semi-infinite chain, see Appendix \ref{2sitesol}. In our case this 
potential is $v^{\a}=W^{\a}+v_{\rm H}$ with Hartree potential  
$v_{\rm H}= 2 U  (\mol-1)\approx 0.7$.

\begin{figure}[t]
\centering
\includegraphics[scale=0.37]{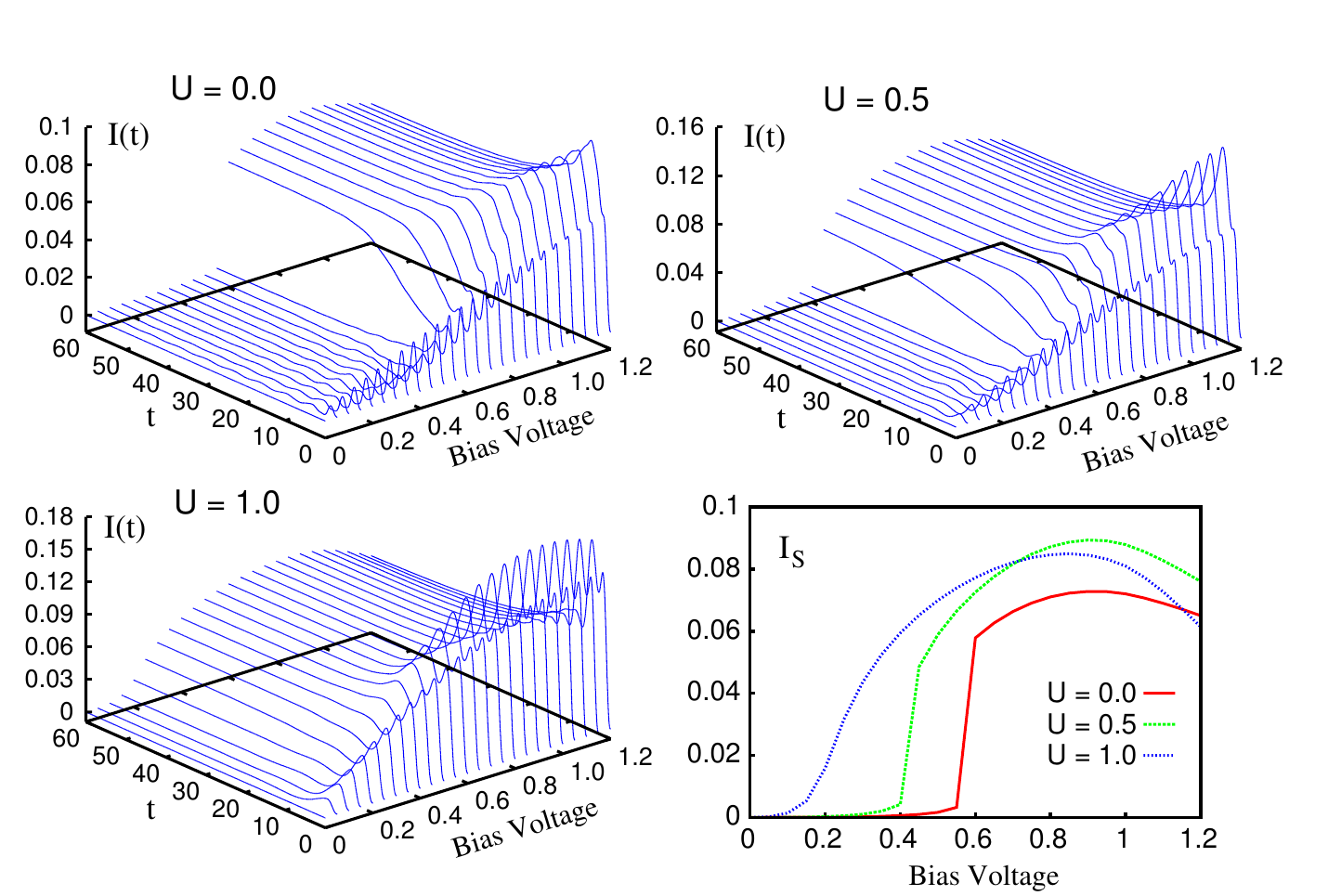}
\caption{2B time-dependent right current for different interaction  
strengths $ U   = 0, 0.5$ (top left and right panels) and $ U  =1.0$ 
(bottom left panel). The $I-V$ curves extracted from the long-time 
limit are displayed in the bottom right panel.}
\label{Fig21}
\end{figure}
\begin{figure}[t]
\centering
\includegraphics[scale=0.37]{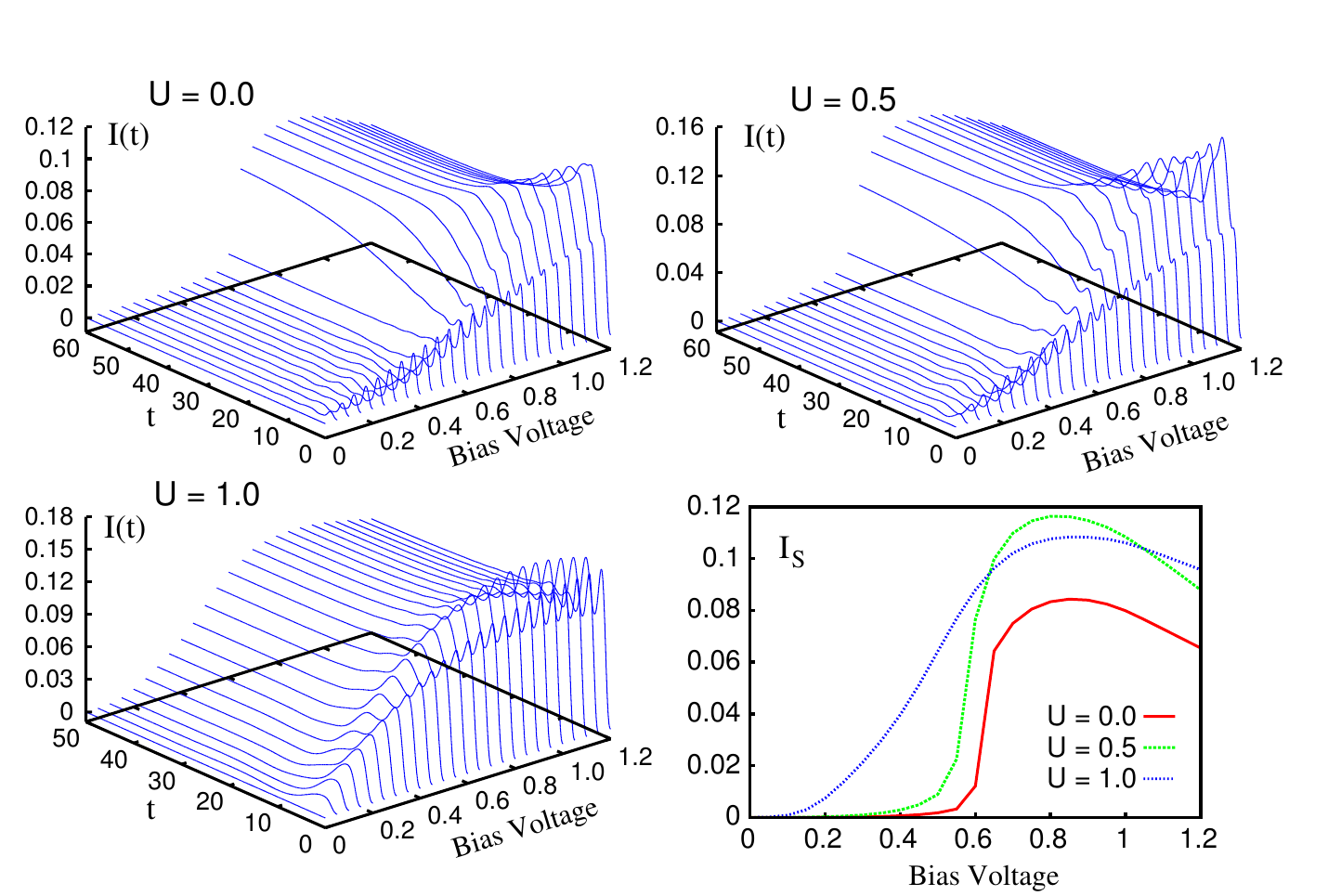}
\caption{GW time-dependent right current for different interaction  
strengths $ U   = 0, 0.5$ (top left and right panels) and $ U  =1.0$ 
(bottom left panel). The $I-V$ curves extracted from the long-time 
limit are displayed in the bottom right panel.}
\label{Fig22}
\end{figure}

The formation of the additional step in the $I-V$ curve
is probably another artifact of the HF approximation.
In Figs.~\ref{Fig21} and ~\ref{Fig22} we show  the transient 
and steady-state currents for $ U  =0, 0.5$ 
and $ U  =1.0$ and bias voltage in the range $[0,1.2]$   
within the  2B and GW approximations. Like for the  HF approximation
the onset of the current is shifted towards smaller bias values when  
$U$ increases. However, this effect is more 
pronounced in the 2B and GW approximations which properly incorporate 
dynamical polarization effects to account for the formation of the image charge. 
Another effect  
of correlations  is to smoothen the onset, in agreement with 
the appearence of a particle-hole shoulder in the spectral function, 
see Section  \ref{scr-rel}.

\begin{figure}[t]
\centering
\includegraphics[trim = 10.5mm 00mm 00mm 0mm, clip, width=9cm]{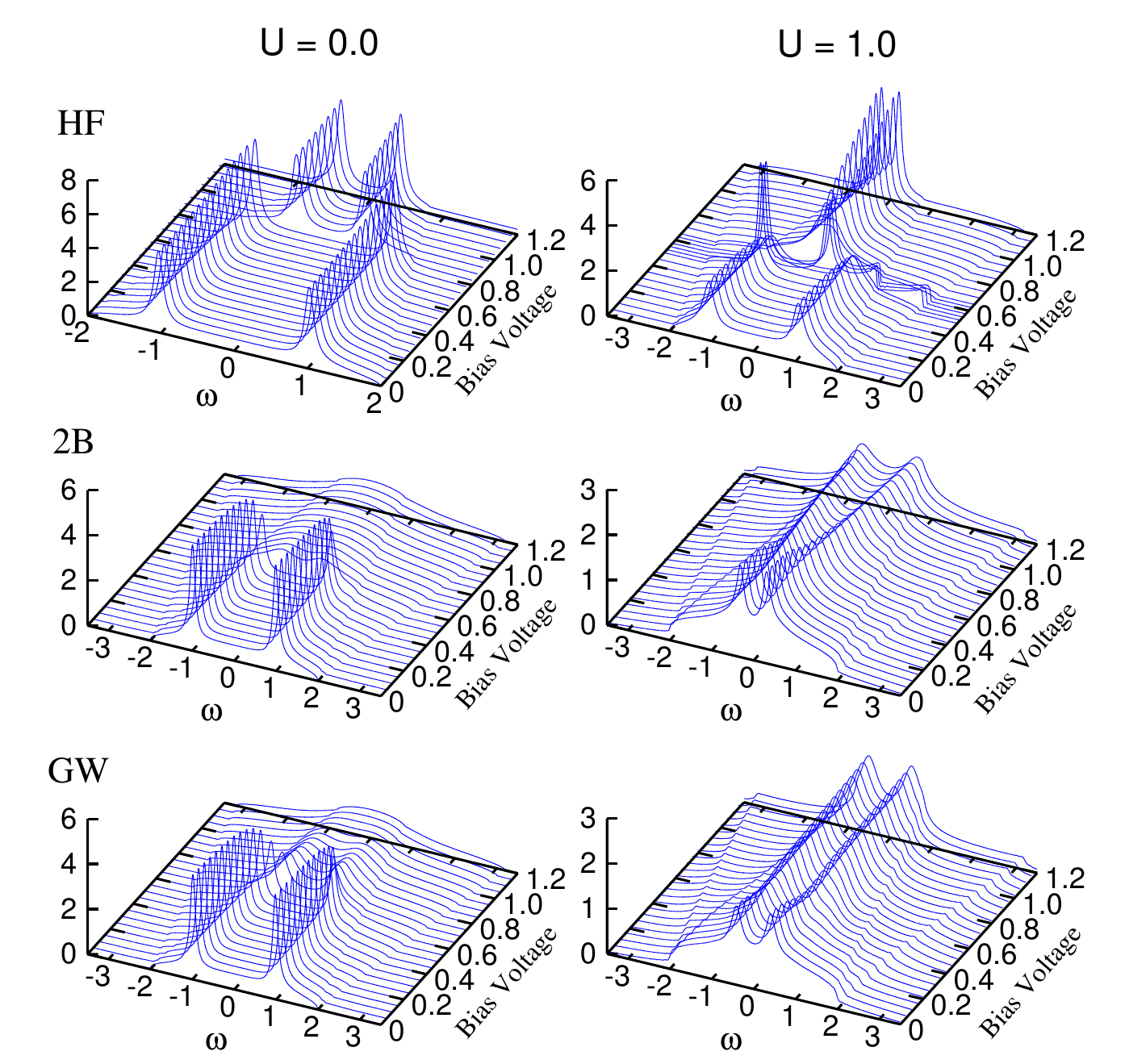}
\caption{HF, 2B and GW spectral functions for different bias 
voltage and lead--molecule interaction $ U  =0.0$ and $ U  =1.0$.} 
\label{Fig23}
\end{figure}

It is important to disentangle the scattering-induced brodeaning 
due to the intramolecular interactions $U_{0}$ and $U_{HL}$ from that due to the 
molecule-lead interaction $U$. In Fig.~\ref{Fig23}  we plot the 
full spectral functions of the interacting region within the HF, 2B and GW approximations for two 
different values of  $ U  =0$ and $ U  =1.0$. 
For $U=0$ and within 2B and GW there is a consistent broadening of the HOMO/LUMO 
peaks when these levels enter the bias window. This effect was 
reported previously in Refs. \onlinecite{kr3.2008} and \onlinecite{Myohanen.2009}.
However, for $U=1$ the 2B and GW spectral functions do not get broader as they enter  
the bias window. The peaks   preserve their shape 
and the HOMO-LUMO  gap starts to open up. 
The molecule-lead interaction  has an  effect opposite to that of the 
intramolecular interaction on the broadening and the many-body shift of 
the spectral peaks.  This is a very important result according to 
which image charge effects in the biased system contribute to lenghten the 
HOMO/LUMO quasi-particle lifetimes and decrease (increase) the ionization 
potential (electron affinity).

\section{Conclusions}
\label{conc}

In conclusion, we provided a thorough analysis of the effects of the 
dynamical formation of image charges at the interfaces between a 
molecule and the metallic leads under non-equilibrium conditions.
The analysis has been carried out within the embedded Kadanoff--Baym 
method using fully self-consistent many-body approximations at the 
HF, 2B and GW level. 
The mean field HF approximation fails to capture the polarization 
effects both in and out of equilibrium. As a consequence, the 
equilibrium molecular levels are not renormalized while out of 
equilibrium the renormalization is solely due to the intramolecular 
interactions. We pointed out that the shortcomings of the HF 
approximation are also at the origin of other unphysical effects. 
There exists a finite range of applied biases for which the molecule 
is artificially charged. This causes a depletion of the electron 
density at the interfaces and prevent the current to increase as the 
bias becomes larger (plateau in the $I-V$ characteristic). Furthermore, for large enough bias and 
molecule-lead interaction the molecular system does not relax in the 
long time limit. We reported the occurrence of the undamped oscillations 
in current and densities. These oscillations correspond  to a charge 
sloshing between the molecular levels and the terminal sites. 

To cure the problems of the mean-field theory we resorted to the 2B 
and GW approximations. In both approximations the self-energy 
contains polarization diagrams which correctly account for the 
screening of the charged molecule and hence are suited to describe the 
formation of image charges. In all situations considered we did not 
observe a plateau in the  $I-V$ characteristic nor the absence of 
relaxation.
An important finding of our analysis is 
that by increasing the molecule-lead interaction the ratio between 
screening time and the relaxation time changes and the screening time is 
primarily determined by the properties of the lead.
As expected, the 2B and GW equilibrium HOMO-LUMO gap closes when increasing the 
molecule-lead interaction. Thus, the onset of 
the current in the $I-V$ characteristic is shifted to lower biases
as compared to a non-interacting or mean-field calculation. 
Another remarkable effect pertains the molecule spectral properties 
as a function of the applied bias. 
In equilibrium the molecule-lead interaction is 
responsible for the reduction of the HOMO-LUMO gap and for a substantial 
redistribution of the spectral weight to the satellites induced 
by the electron correlations. Increasing the bias the situation 
changes. For zero molecule-lead interaction the HOMO and LUMO peaks 
near each other and considerably broaden when they enter the bias 
window. The effect of the molecule-lead interaction is to keep the 
spectral peaks sharp and to open the HOMO-LUMO gap. This effect is 
therefore exactly the opposite of that generated by the 
intramolecular interactions.
All this phenomenology clearly shows the 
importance  of a proper description of electron correlations in 
time-dependent and steady-state quantum transport.

\acknowledgments
The authors want to thank Adrian Stan and Anna-Maija Uimonen for fruitful discussions.
We also like to thank the Academy of Finland and the
Vilho, Yrj\"o and Kalle V\"ais\"al\"a foundation for financial support and 
CSC IT center for providing resources for scientific computing. 

\appendix

\section{Other exact results of the Image Charge Model in the uncontacted case}
\label{2sitesol}

\begin{figure}[t]
\centering
\includegraphics[scale=0.5]{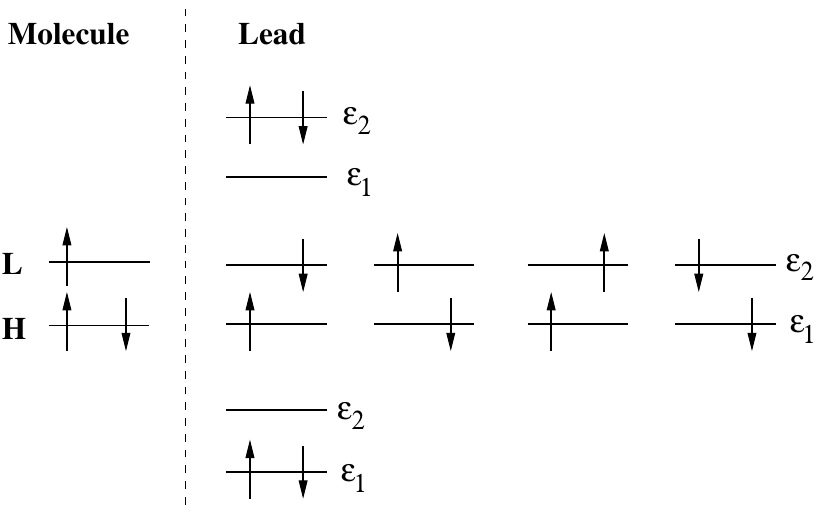}
\caption{Electronic configuration for a two-site lead. 
For an extra electron on the molecule
there are three energy eigenspaces for the lead.
Note that the parallel-spin electron states do not contribute in the 
response properties since the bias
preserves the spin orientation.} 
\label{Fig24}
\end{figure}

In this Appendix we derive some simple analytic result for the ICM 
with $\l^{l}=\l^{r}=0$, $U^{l}=0$ and $U^{r}=U$. We will show that 
the main qualitative features of the system in equilibrium can be 
captured already by considering leads of finite length. Let us 
consider the molecule with an extra electron on the LUMO level and a 
right lead with only two sites. The extra electron induces an 
impurity-like potential $U$ on the terminal site and the 
single-particle eigenvalues of the lead Hamiltonian are then given  
by 
$\varepsilon_{1,2} = \mp \sqrt{( U/2)^2 + b^2}$. 
Let us denote by $M={\rm GS}^{+}$ the molecular configuration with 
the extra electron. At half-filling the right lead has two electrons and the 
eigenstates $|{\rm GS}^{+} s\ket$ of Eq.~(\ref{nonintchain}) are 
displayed in Fig.~\ref{Fig24}. Their energy is  
$\mathcal{E}_0 = 2\varepsilon_1$, $\mathcal{E}_1 = \varepsilon_1 + 
\varepsilon_2$ and $\mathcal{E}_2 = 2\varepsilon_2$.
In a similar manner we can calculate the lead eigenenergies for the 
molecule with an electron less. The resulting ionization potential 
and electron affinity are $I=E_{4} - E_{3} = \epsilon_H + U_0 - 2|b| + 2\sqrt{( U/2)^2 + b^2}$
and $A=E_{5}-E_{4} = \epsilon_L + 2U_{HL} + 2|b| - 
2\sqrt{( U  /2)^2 + b^2}$, see Eqs. (\ref{aff}) and (\ref{ion}).
These energies correspond to the renormalized energies of the HOMO 
and LUMO level.
We thus see that increasing the Coulomb interaction $ U  $ the HOMO 
and LUMO levels approach each other,  in agreement with the general 
result of Section \ref{exsol}.

The density at the terminal site is unity if the molecule is charge 
neutral. However, since the molecule with one electron more/less 
induces an impurity-like potential $\mp U$, the terminal site density in this 
case changes according to 
\be
n_1(u) = \int_{-\infty}^{\mu}A_{11}(\omega,u) \ud \omega,
\ee
where $A_{11}(\omega,u) =  -\frac{1}{\pi}\Im[G_{11}^R(\omega,u)]$ is the spectral function
projected on the terminal site with an impurity-like potential $u=\mp U$. 
The Green's function can be  calculated explicitly from the Dyson 
equation and reads
\be\label{ExactG}
G_{11}^R(\omega, U  ) = G_{11}^{0,R}(\omega)/(1- U   G_{11}^{0,R}(\omega)).
\ee
Here $G_{11}^{0,R}$ is the unperturbed retarded Green's function of the 
semi-infinite lead and it reads
\be\label{Gzero}
G_{11}^{0,R}(\omega) = \frac{1}{2b^2}  \left\{
 \begin{array}{c }
 (\omega - \mathrm{sgn} (\omega) \sqrt{\omega^2-4b^2})  \quad (|\omega| > 2 |b|) \\
 (\omega - i \sqrt{4b^2- \omega^2} ) \quad (|\omega| < 2 |b|)
 \end{array} \right.
\ee
\begin{figure}[t]
\centering
\includegraphics[scale=0.35]{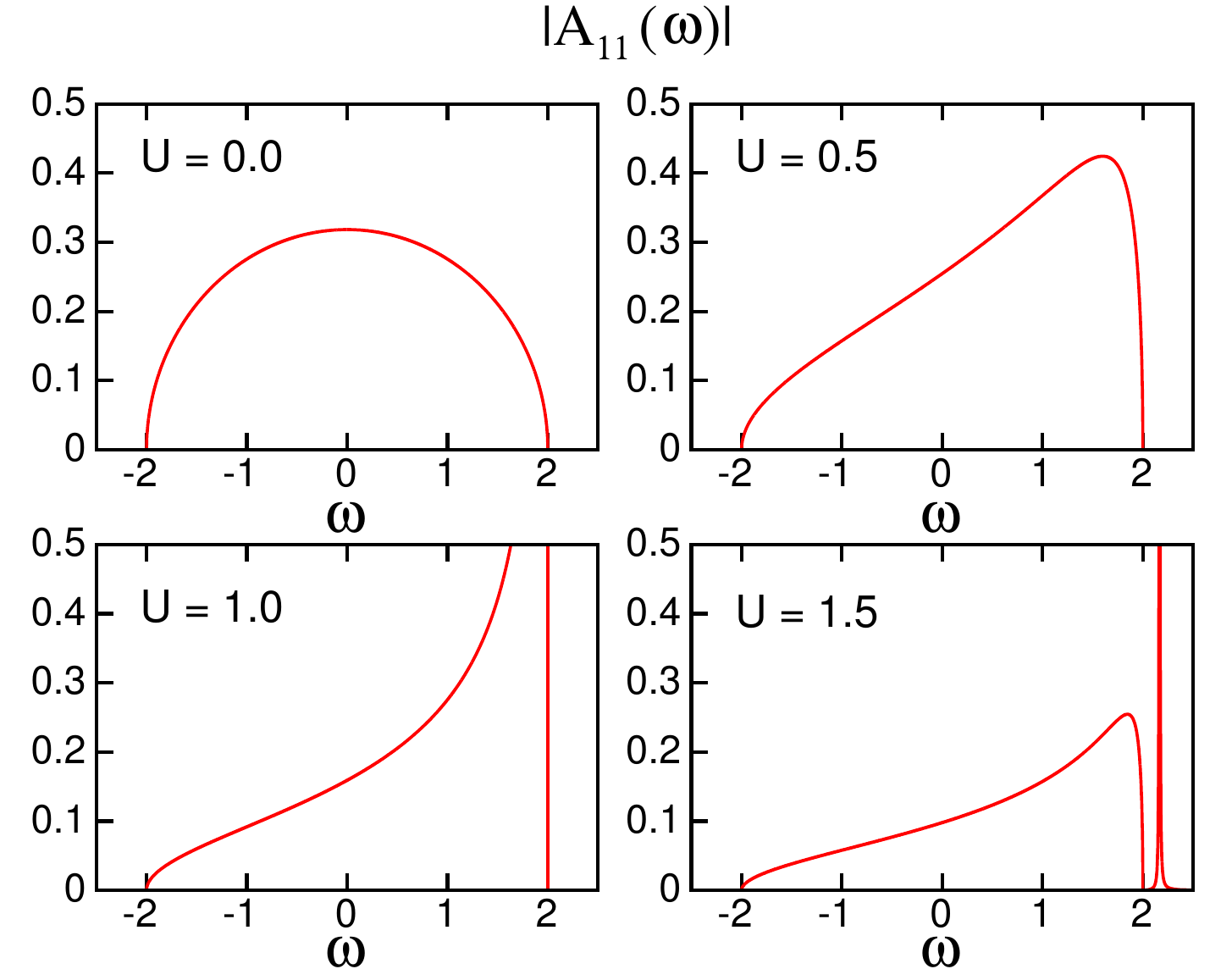}
\caption{Formation of the split-off state (sharp peak in the 
bottom right panel) as $U$ increases.  
In all plots  $b=-1.0$.} 
\label{Fig25}
\end{figure}

\noindent
If $ |U| $ exceeds   the lead hopping $b$ a split-off state 
appears outside the energy continuum.  This is illustrated in 
Fig. \ref{Fig25} where  
we plot the lead spectral function $A_{11}(\omega, U  )$ for $ U  =0, 
\ 0.5, \ 1.0,\ 1.5$. This  
split-off state appears as a pole in the Green function of Eq. (\ref{ExactG}) with the energy 
\be
\epsilon( U  ) = b\left[\frac{1+\left(\frac{ U  
}{b}\right)^2}{\left(\frac{ U  }{b}\right)}\right] 
\ee
Comparing the spectral structure of Fig. \ref{Fig25} with 
that of Fig. \ref{Fig20} we conclude that the extra step in 
the HF $I-V$ curve is due to the formation of a split-off state which 
prevents the current to incease as the bias becomes larger.

\section{Density response function}
\label{DRF}

We here calculate the density response function projected onto the terminal site of a 
semi-infinite TB chain relevant for the discussion of 
Section \ref{exsol}. For chains with $N_{\rm ch}$ sites the single-particle 
eigenfunctions and eigenenergies of the system  
are  $\psi_k(i) = 
(-1)^{i+1}\sqrt{\frac{2}{N_{\rm ch}+1}}\sin (\phi_k i)$ and 
 $\epsilon_k = 
-2b\cos(\phi_k)$, where $\phi_k = \frac{k\pi}{N_{\rm ch}+1}$, 
$k=1...N_{\rm ch}$. By definition, the (retarded) density response function 
$\chi_{ij}(\w)$ with site coordinates $(i,j)$ reads
\begin{equation}
\chi_{ij}(\omega) = 2\sum_{kl}(f_k-f_l) 
\frac{\psi_k^*(i)\psi_l(i)\psi_k(j)
\psi_l^*(j)}{\omega - (\epsilon_l-\epsilon_k) + \mathrm{i}\eta}  
\end{equation}
where, for zero temperature, $f_k = \theta(\m - \e_{k})$ are the single-particle 
occupations and $\eta$ is an infinitesimally small positive constant. 
Inserting the explicit form of the eigenfunctions and eigenvalues, 
changing the variables to $\bar{k} = 
k\pi/(N_{\rm ch}+1)$ and  taking the $N_{\rm ch}\rightarrow \infty$ limit, we get for 
the $i=j=1$ component 
\be
\chi_{11}(\omega)= 
\frac{8}{\pi^2}\int_0^{\pi}\ud\bar{k}\int_0^{\pi}\ud\bar{l} \,
\frac{(f_{\bar{k}}-f_{\bar{l}})\sin^2\bar{k}\sin^2\bar{l}}{\omega 
- 2b(\cos\bar{k} - \cos\bar{l} ) + \mathrm{i}\eta}, 
\nonumber
\ee
where for the half-filled system here considered $f_{\bar{k}} = 
\theta(\frac{\pi}{2} - \bar{k})$. This  
expression can be simplified further by changing the variables  
to $x = \cos\bar{l}$ and $y = \cos\bar{k}$. The  integral containing 
$f_{\bar{k}}$ becomes 
\beq
\chi_{11}^{(1)}(\omega) &=& \frac{8}{\pi^2}
\int_0^1\ud y\int_{-1}^{1}\ud x 
\frac{\sqrt{1-y^2}\sqrt{1-x^2}}{\omega-2b(y-x)+i\eta} \nonumber\\ 
&=& \Lambda^{(1)}(\omega) - \mathrm{i}\Delta^{(1)}(\omega),
\eeq
where
$\Lambda^{(1)}(\omega) = \frac{1}{\p}\mathcal{P}\int d\w' 
\frac{\Delta^{(1)}(\omega')}{\w-\w'}$ is the real part and
\beq
\Delta^{(1)}(\omega) &=& \frac{4}{\pi|b|}\int_0^1 \ud y 
\sqrt{1-y^2}\sqrt{1-\left[y-\omega/(2b)\right]^2} \times \nonumber 
\\  
&&\theta(y-\omega/(2b)+1)\theta(1-\left(y-\omega/(2b)\right)),
\eeq
is the imaginary part. Similarly one obtains the  integral 
$\chi^{(2)}_{11}$ containing $f_{\bar{l}}$. The sum 
$\chi_{11}=\chi_{11}^{(1)}+\chi_{11}^{(2)}$
can now easily be calculated numerically.

\section{Explaining the level broadening in the HF approximation}
\label{renormalization}

In this Appendix we show that the molecule-lead interaction in the 
presence of a finite hybridization renormalizes the embedding 
self-energy already in the HF approximation, thus explaining the 
broadening of the HF peaks in Fig. \ref{Fig8}.
Let us denote simply by $G$ and $\S$ the retarded components of the 
Green function and self-energy respectively. For simplicity we take 
$\l^{l}=U^{l}=0$ and $\l^{r}=\l$, $U^{r}=U$ and we denote by $1$ the terminal site of the 
right lead. We start from the Dyson equation for $G(\w)$ 
\be
(\omega -h - \S^{\mathrm{HF}})G(\omega) = 1
\ee
where, in accordance with the notation of Section \ref{mb-treatment}, 
$h$ is the Hamiltonian in the one-particle Hilbert space and has the 
structure
\be
h=\left(\begin{array}{ccc}
\e_{H} & 0 & h_{H,r} \\
0 & \e_{L} & h_{L,r} \\
h_{r,H} & h_{r,L} & h_{r,r}
\end{array}
\right).
\ee
Here $h_{r,r}$ is the tridiagonal matrix which describes the right 
lead with matrix elements $b$ on the upper and lower diagonal and 
zero otherwise, while $h_{i,r}$, with $i=H,L$, is the rectangular 
matrix whose only non-vanishing entry is $(h_{i,r})_{i,1}=\l$.
Projecting the Dyson equation onto $HH$ and $r,H$ we find
\beq
(\omega - \e_{H} - \S_{HH}^{\mathrm{HF}})G_{HH}(\omega) 
&=&1 +  [h_{H,r} + 
\S_{H,r}^{\mathrm{HF}}]G_{r,H}(\omega)\nonumber\\ 
(\omega  - h_{r,r}-\S_{r,r}^{\mathrm{HF}})G_{r,H}(\omega) 
&=& [h_{r,H} + 
\S_{r,H}^{\mathrm{HF}}]G_{HH}(\omega).\nonumber 
\eeq
Solving the second equation for $G_{r,H}$ and inserting the result in 
the first equation we obtain the following solution for $G_{HH}$
\be
G_{HH}(\w)=\frac{1}{\w-\e_{H}-\S_{HH}}
\ee
with 
\be
\S_{HH}=\S^{\rm 
HF}_{HH}+(\l+\S^{\rm HF}_{H1})\tilde{G}_{11}(\w)(\l+\S^{\rm HF}_{1H}).
\label{shhtot}
\ee
In the above equation $\tilde{G}_{11}$ is the (1,1) matrix element of the Green's function of the 
uncontacted system ($\l=0$) with the same HF self-energy, i.e., 
$\tilde{G}_{r,r}=1/(\w-h_{r,r}-\S^{\rm HF}_{r,r})$. Note that the 
only non-vanishing entry of the self-energy in the lead is 
$(\S^{\rm HF}_{r,r})_{11}=\S^{\rm HF}_{11}$.
Next we observe that the nonlocal HF self-energy can  be written as 
\be
\S_{1H}^{\mathrm{HF}} =  iU  
\int_{-\inf}^{\m}\frac{d\omega}{2\pi}\left(-2i{\rm Im}[G_{1H}(\omega)]\right),
\label{hfs1h}
\ee
and similarly for $\S_{1H}^{\mathrm{HF}}$. From the projected Dyson 
equation we have
\be
G_{1H}(\w)=\tilde{G}_{11}(\w)(\l+\S^{\rm HF}_{1H})G_{HH}(\w).
\label{g1h}
\ee
For the equilibrium system we can always choose the HF orbital to be 
real valued and therefore ${\rm Im}[\S^{\rm HF}_{1H}]=0$. Then, 
inserting Eq. (\ref{g1h}) into Eq. (\ref{hfs1h}) and solving for 
$\S_{1H}^{\mathrm{HF}}$ we find
\be
\S_{1H}^{\mathrm{HF}}=\l\frac{UC}{1-UC}
\ee
where 
\be
C=2\int_{-\inf}^{\m}\frac{d\omega}{2\pi} {\rm 
Im}\left[\tilde{G}_{11}(\w)G_{HH}(\w)\right]
\label{Cconst}
\ee
This result together with its analogous for $\S_{H1}^{\mathrm{HF}}$ 
allows us to cast the self-energy in Eq. (\ref{shhtot}) in the form
\be
\S_{HH}=\S^{\rm HF}_{HH}+\left(\frac{1}{1-UC}\right)^{2}\S^{\rm 
em}_{HH}(\w)
\ee
\begin{figure}[t]
\centering
\includegraphics[scale=0.55]{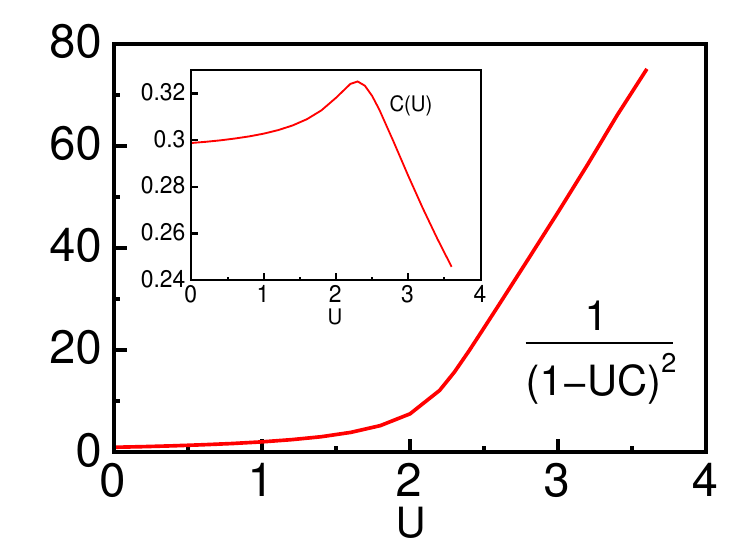}
\caption{ The value of the renormalization constant as a function of $U$
for $b=-1.0$ and $\lambda=-0.2$. The inset shows the dependence of the
factor $C$ as a function of $U$.} 
\label{Fig26}
\end{figure}
where $\S^{\rm em}_{HH}(\w)=\l^{2}\tilde{G}_{11}(\w)$ is the 
embedding self-energy of the non-interacting system. Thus, the 
molecule-lead interaction renormalizes the embedding self-energy and 
increases the  broadening of the HF spectral peaks. 
The value of the constant $C$ in Eq.(\ref{Cconst}) can be
determined numerically. 
In Fig. \ref{Fig26} we display $(1-CU)^{-2}$ and $C$ (inset)
as a function of $U$. We see that $C$ is roughly constant for small $U$.

\end{document}